\documentclass[11pt,twoside]{article}
\usepackage{fancyhdr}
\usepackage{amsfonts,epsfig,graphicx}
\usepackage{afterpage}
\usepackage{amsmath,amssymb,amsthm} 
\usepackage{fullpage}
\usepackage[T1]{fontenc} 
\usepackage{epsf} 
\usepackage{graphics} 
\usepackage{enumitem}
\usepackage{amsfonts,amsmath}
\usepackage[linesnumbered,ruled,vlined]{algorithm2e}
\usepackage{algorithmic}
\usepackage[]{natbib} 
\usepackage{psfrag,xspace}
\usepackage{color,etoolbox}
\usepackage[font=small,labelfont=bf]{caption}
\usepackage{subfigure} 
\usepackage{hyperref}[] 
\hypersetup{
colorlinks = true,
linkcolor = blue, 
filecolor = magenta,
urlcolor = blue, 
citecolor = blue 
}

\usepackage{booktabs}

\newtheorem{theorem}{Theorem}[section]
\newtheorem{definition}{Definition}[section]

\newtheorem{proposition}{Proposition}[section]
\newtheorem{corollary}{Corollary}[section]
\newtheorem{remark}{Remark}[section]
\newtheorem{example}{Example}[section]

\newcommand{\mP}{\mathbb{P}}
\newcommand{\mE}{\mathbb{E}}

\newcommand{\mT}{\mathcal{T}} 	
\newcommand{\mA}{\mathcal{A}}

\newcommand{\convP}{\overset{p}{\longrightarrow}}
\newcommand{\convD}{\overset{d}{\longrightarrow}}

\begin{document}

\title{}

\begin{center} {\Large{\bf{Global and Local Two-Sample Tests via Regression}}}
	
	\vspace*{.3in}
	
	{\large{
			\begin{tabular}{ccccc}
				Ilmun Kim$^\dagger$&&Ann B. Lee$^\dagger$ &&Jing Lei $^\dagger$ \\
			\end{tabular}
			
			\vspace*{.1in}
			
			\begin{tabular}{ccc}
				Department of Statistics and Data Science$^{\dagger}$ \\
			\end{tabular}
			
			\begin{tabular}{c}
				Carnegie Mellon University \\
				Pittsburgh, PA 15213
			\end{tabular}
	}}
	
	\vspace*{.2in}
	
	\begin{abstract}
		Two-sample testing is a fundamental problem in statistics. Despite its long history, there has been renewed interest in this problem with the advent of high-dimensional and complex data. Specifically, in the machine learning literature, there have been recent methodological developments such as classification accuracy tests. The goal of this work is to present a regression approach to comparing multivariate distributions of complex data. Depending on the chosen regression model, our framework can efficiently handle different types of variables and various structures in the data, with competitive power under many practical scenarios. Whereas previous work has been largely limited to global tests which conceal much of the local information, our approach naturally leads to a local two-sample testing framework in which we identify local differences between multivariate distributions with statistical confidence. We demonstrate the efficacy of our approach both theoretically and empirically, under some well-known parametric and nonparametric regression methods. Our proposed methods are applied to simulated data as well as a challenging astronomy data set to assess their practical usefulness.
	\end{abstract}
\end{center}

\vskip 1em

\section{Introduction} \label{Sec: Introduction}
Given two distributions $P_0$ and $P_1$ on $\mathbb{R}^D$, the global two-sample problem is concerned with testing $H_0: P_0 = P_1$ versus $H_1: P_0\neq P_1$, based on independent random samples from each distribution. This fundamental problem has a long history in statistics and has been well-studied in a classical setting \citep[see, e.g.,][]{thas2010comparing}. Recently, however, there has been renewed interest in this field as modern data we encounter have become more complex and diverse. Traditional approaches, which focus on low-dimensional and Euclidean data, often fail or are not easily generalizable to high-dimensional and non-Euclidean data. Additionally, some recent developments in high-dimensional two-sample testing are limited to simple alternatives such as location and scale differences \citep[see,][for a recent review]{hu2016review}. In this context, there is a need to develop a new tool for the two-sample problem that can efficiently handle complex data and can detect differences beyond location and scale alternatives.

When the null hypothesis of the global two-sample test is rejected, it is often valuable (for e.g. scientific discovery, calibration of simulation models, and so on) 
to further explore {\em how} the two distributions are different. Specifically, as a follow-up study to the global test, one might wish to identify locally significant regions where the two distributions differ. This topic, 
which we refer to as the {\em local two-sample problem}, has been studied by \cite{duong2013local} who uses kernel density estimators to identify local differences between two density functions. However, 
the kernel density approach may perform poorly when distributions are not in a low-dimensional Euclidean space, and hence another tool is needed for more challenging settings.

The goal of this work is to develop a general framework for both global and local two-sample problems that overcomes the aforementioned challenges. Specifically, we aim to design a two-sample test that can efficiently handle different types of variables (e.g.~mixed data types) and various structure (e.g.~manifold, irrelevant covariates) in the data. Consequently, the resulting test can have substantial power for a variety of challenging alternatives. We achieve our goal by connecting the two-sample problem to a regression problem as follows. Let $f_0$ and $f_1$ be density functions of $P_0$ and $P_1$ with respect to a common dominating measure. We view $f_0$ and $f_1$ as conditional densities $f(x|Y=0)$ and $f(x|Y=1)$ by introducing an indicator random variable $Y \in \{0,1\}$. Then by Bayes' theorem, the hypothesis $H_0: f_0(x)= f_1(x)$ for all $x \in S= \{x \in \mathbb{R}^D: f_0(x) + f_1(x) > 0\}$ can be verified to be equivalent to the hypothesis that involves the regression function:
\begin{align} \label{Eq: hypothesis_statement}
H_0: ~ \mP(Y=1 |X=x) = \mP(Y=1), \ \ \text{for all } x \in S.
\end{align} 
We state the corresponding global and local alternative hypotheses as
\begin{align*}
H_1 : ~& \mP(Y=1 |X=x) \neq  \mP(Y=1), \ \ \text{for some~} x \in S,  \text{ and} \\[.5em]
H_1(x) : ~& \mP(Y=1 |X=x) \neq  \mP(Y=1), \ \ \text{at fixed~} x \in S,
\end{align*}
respectively. 

Motivated by the above reformulation, we propose a testing procedure that measures an empirical distance between the regression function $\mP(Y=1|X=x)$ and the class probability $\mP(Y=1)$. We refer to this approach as {\em the regression test}. Depending on the choice of regression method, the regression test can adapt to nontraditional data settings. As we shall see, the power of the test is closely related to the mean square error of the chosen regression estimator. In addition, by choosing a nonparametric regression method, the global regression test can be sensitive to general alternatives beyond location and scale differences. 
We will demonstrate the benefits of the regression test with both theoretical and empirical results. 

\subsection{Motivating Example} \label{Section: Motivating Example}
We motivate our approach by comparing multivariate distributions of galaxy morphologies, but the proposed framework benefit other areas of science and technology as well (involving, e.g., outlier detection, calibration of simulation models, and comparison of cases and controls).
A galaxy's morphology is the organization of a galaxy's light, as projected into our line of sight and observed at a particular wavelength as a pixelated image. Morphological studies are key to understanding the evolutionary history of galaxies and to constraining theories of the Universe; see e.g. \cite{conselice2014evolution} for a review. So far astronomers have only been able to study one or two morphological statistics (or projections of these) at a time
instead of an entire ensemble. The reason is  
a lack of tools for effectively comparing and jointly analyzing multivariate or high-dimensional data in their native spaces. A global hypothesis test  with a binary reject yes/no answer is also not informative enough to explain how two distributions are different in a multivariate feature space. 

\begin{figure}[!t]
	\begin{center}
		\includegraphics[width=\textwidth]{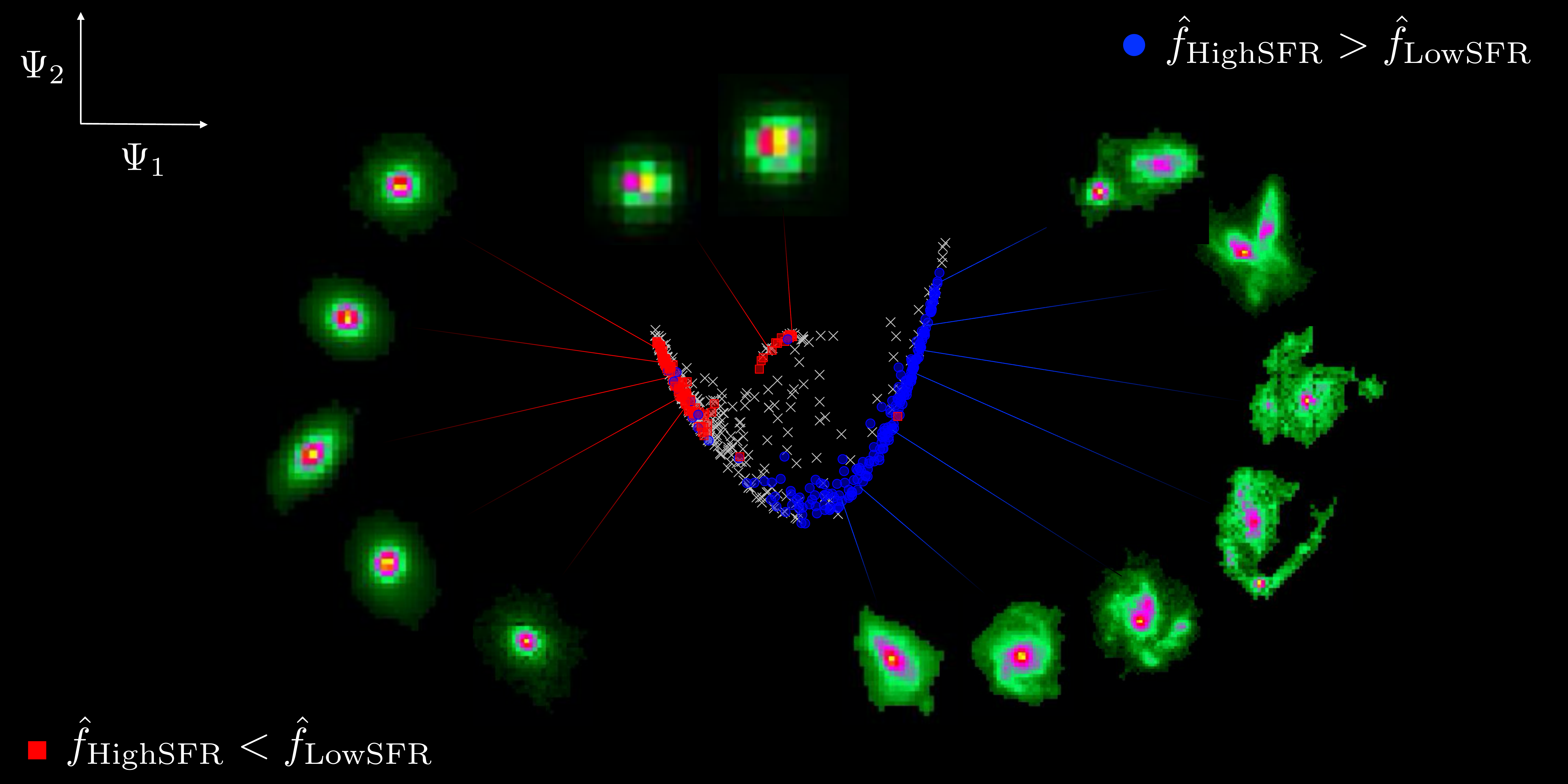}
		\caption{Result of local two-sample test of differences between high- and low-SFR galaxies in a seven-dimensional morphology space. The red squares indicate regions where the density of low-star-forming galaxies are significantly higher, and the blue circles indicate regions in morphology space that are dominated by high-star-forming galaxies; the gray crosses represent insignificant test points. The galaxies are embedded in a two-dimensional diffusion space for visualization purposes only (see Appendix~\ref{Sec: Diffusion Map} for details); $\Psi_1$ and $\Psi_2$ here denote the first two coordinates.} \label{Figure: Diffusion Map}
	\end{center}
\end{figure}

We illustrate the efficacy of the proposed global and local testing framework on the morphology statistics of two galaxy populations with high and low star-formation rate (SFR), respectively. 
The challenge here is not only that the problem involves multivariate data, but also that some of the morphological statistics are mixed discrete and continuous type with heavy outliers. 
We efficiently handle this issue by building on the success of random forests regression. The visualized local two-sample result is shown in Figure~\ref{Figure: Diffusion Map} and the details of the analysis can be found in Section~\ref{Section: Application to Astronomy Data}.

\subsection{Related Work}
In recent years, several attempts have been made to connect binary classification with two-sample testing. The main idea of this approach is to check whether the accuracy of a binary classifier is better than chance level and reject the null if the difference is significant. Such an approach, referred to as an accuracy or classification test, was conceptualized by \cite{friedman2003multivariate} and has since been investigated by several authors \citep[][]{ojala2010permutation, olivetti2015statistical, ramdas2016classification, rosenblatt2016better, gagnon2016classification, lopez2016revisiting,hediger2019use}. 
In the same manner as our regression framework, a key strength of the accuracy test is that it offers a flexible way for the two-sample problem as it can utilize any existing classification procedure in the literature. However, the classification accuracy framework is not easily converted to a local two-sample test. In addition, many classifiers are estimated by dichotomizing regression estimators and the discrete nature of such classifiers may result in a less powerful test (see Section~\ref{Section: A Comparison between Regression and Classification Accuracy Tests} and other simulation results). 

For the global two-sample test, our framework can be viewed as an instance of goodness-of-fit testing for regression models \citep[e.g.][for a review]{gonzalez2013updated}. There is a substantive literature on this topic including \cite{hardle1993comparing}, \cite{weihrather1993testing}, \cite{gonzalez1993testing}, \cite{zheng1996consistent}, \cite{zhang2004power}, \cite{hart2013nonparametric} and among others. This line of work typically concentrates on comparing differences between parametric (e.g.~linear regression) and nonparametric (e.g.~kernel regression) fits from an asymptotic point of view. For example, \cite{hardle1993comparing} consider the squared deviation between a parametric regression estimator and a kernel estimator. They show that their test statistic converges to a normal distribution under the null hypothesis and justify the use of the wild bootstrap procedure. However, this type of asymptotic approach is challenging to analyze beyond kernel-type estimators and often requires strong technical assumptions. 
In contrast, our framework is designed to compare any type of regression estimators with a specific constant fit by building upon the permutation principle. Hence the resulting test is valid in any finite sample sizes. Moreover we present a unified framework of studying the power of the regression test by taking advantage of existing results on the estimation error.

For the local two-sample test, our approach has similarities to independent work by \cite{cazais2015beyond} who estimate the Kullback-Leibler divergence between $\mP(Y=1|X=x)$ and $\mP(Y=1)$. Our procedure, however, identifies locally significant areas with statistical confidence whereas \cite{cazais2015beyond} graphically decide a threshold for the significance.

\subsection{Overview of this paper}
We outline the paper as follows: In Section~\ref{Section: Framework}, we introduce the proposed metrics, test statistics and algorithms for the global and local regression tests. In Section~\ref{Section: Global Two-Sample Tests via Regression}, we study theoretical properties of the global regression test. We begin by considering a simple scenario where two populations only differ in their means in Section~\ref{Section: Fisher's Linear Discriminant Analysis}. In this scenario, we show that the regression test based on Fisher's linear discriminant analysis (LDA) achieves the same local optimality as the Hotelling's $T^2$ test. Moving on to general regression settings in Section~\ref{Section: The MISE and Testing Error for Global Regression}, we establish a connection between the testing error of the global regression test and the mean integrated square error (MISE) of the regression estimator. In Section~\ref{Section: Local Two-Sample Tests via Regression}, we turn to the local two-sample problem and investigate general properties of the local regression tests.  In Section~\ref{Section: The MSE and Testing Error for Local Regression}, we describe the testing error of the local regression test in terms of the mean square error (MSE) of the regression estimator. We further establish an optimality of the local regression tests over the Lipschitz class from a minimax point of view in Section~\ref{Section: Minimax Optimality over the Lipschitz Class}. When data have intrinsic dimension, we show that the performance of the local regression tests based on kNN or kernel regression only depends on intrinsic dimension in Section~\ref{Section: Adaptation to Intrinsic Dimension}. Section~\ref{Section: Limiting Distribution of Local Permutation Test Statistics} studies the limiting distribution of the local permutation statistic to avoid a high computational cost from permutations for large sample size. In Section~\ref{Section: Simulations}, simulation studies are provided to illustrate finite sample performance of the global and local regression tests. In Section~\ref{Section: Application to Astronomy Data}, we apply the proposed approach to a problem in astronomy and demonstrate its efficacy. All the proofs are deferred to Appendix~\ref{Section: Proofs}.

\vskip .8em

\paragraph{Notation.} Throughout this paper, we denote the class probabilities $\mP(Y=0)$ and $\mP(Y=1)$ by $\pi_0$ and $\pi_1$, respectively, and write the joint distribution of $(X,Y)$ by $\pi_0 [P_0 \times \delta_0] + \pi_1 [P_1 \times \delta_1]$ where $\delta_k$ denotes the point mass at $k$ for $k=0,1$. We denote the corresponding conditional probability $\mP(Y=1|X=x)$ by $m(x)$, which can be explicitly written as
	\begin{align*}
	m(x) = \frac{\pi_1 f_1(x)}{\pi_1 f_1(x) + \pi_0 f_0(x)}.
	\end{align*}
	We use $P_X(\cdot)$ to denote the marginal probability measure of $X$ and $|| Z ||_2$ denotes the Euclidean norm of a vector $Z \in \mathbb{R}^D$. The symbols $\convP$ and $\convD$ stand for convergence in probability and in distribution, respectively. We use $a_n \lesssim b_n$ if there exists $C>0$ such that $a_n \leq C b_n$ for all $n$. Similarly, $a_n \asymp b_n$ if there exist constants $C,C^\prime >0$ such that $C \leq |a_n/b_n| \leq C^\prime$ for all $n$. As convention, the acronym \emph{i.i.d.} is used to represent independent and identically distributed.


\section{Framework} \label{Section: Framework}

\subsection{Metrics} \label{Section: Metrics}
A common metric for comparing two distributions is the difference between two density functions $f_0(x)$ and $f_1(x)$; this metric has been used for global and local two-sample testing by \cite{anderson1994two} and \cite{duong2013local}. Another natural metric, suggested for global two-sample testing by \cite{keziou2005test}, \cite{fokianos2008comparing} and \cite{sugiyama2011least}, is the density ratio $f_1(x)/f_0(x)$. Although 
both the density difference and density ratio metrics are intuitive, there are several weaknesses associated with each of them. For example, the estimation of a density difference is largely limited to kernel density estimators, which are sensitive to the curse of dimensionality. The density ratio, on the other hand, could potentially be estimated  using various regression methods thanks to the following reformulation:
\begin{align*}
\frac{f_1(x)}{f_0(x)} = \frac{\pi_0}{\pi_1} \frac{m(x)}{1-m(x)},
\end{align*}
\citep[see, e.g.,][]{qin1997goodness}. The main weakness of the ratio approach, however, is that the ratio is highly sensitive to the tail behavior of distributions, and it is not even well defined when $m(x)=1$. 
To overcome these limitations, 
we propose an alternative approach which instead compares the regression function with the class probability. More specifically, we consider 
\begin{align} \label{eq: population quantity}
\mathcal{T}_{global} = \int_S \{ m(x) - \pi_1 \}^2 dP_X(x),  \quad  \mathcal{T}_{local}(x) = \{ m(x) - \pi_1 \}^2
\end{align}
as global and local measures of the discrepancy between two distributions where we assume that $\pi_1$ is a fixed constant within $0<\pi_1 <1$ throughout this paper. By construction, both $\mathcal{T}_{global}$ and $\mathcal{T}_{local}(x)$ are bounded between zero and one. More importantly, we can take advantage of numerous existing regression methods \citep[see, e.g.,][for popular methods and descriptions]{friedman2009elements} when estimating $m(x)$. Hence, our approach maintains the flexibility of the density ratio approach while avoiding the problem of ill-defined quantities.

\subsection{Test Statistics and Algorithms} \label{Sec: Test Statistics and Algorithm}
Suppose we observe $n$ pairs of samples $\{(X_i,Y_i)\}_{i=1}^n$, where $X_i \in \mathbb{R}^D$ and $Y_i \in \{0,1\}$. Let $\widehat{m}(x)$ be an estimate of $m(x)$ based on the samples, and $\widehat{\pi}_1 =  \frac{1}{n} \sum_{i=1}^n I(Y_i=1)$. Then by plugging these statistics into (\ref{eq: population quantity}), we define our global and local test statistics as
\begin{align} \label{Eq: Test Statistic}
\widehat{\mT}_{global} = \frac{1}{n} \sum_{i=1}^n \{ \widehat{m}(X_i) - \widehat{\pi}_1 \}^2, \quad \widehat{\mT}_{local}(x) =  \{ \widehat{m}(x) - \widehat{\pi}_1 \}^2.
\end{align} 
The null distributions of the proposed test statistics are typically unknown, and they depend on the choice of regression method as well as the distribution of the data.  Hence, to keep our framework as general as possible, we use a permutation procedure to set a critical value that yields a valid level $\alpha$ test for any given regression estimator under any sampling scheme given in Section~\ref{Section: Sampling Schemes}. The proposed permutation framework for global and local two-sample testing are summarized in Algorithm~\ref{Alg: Global Two-Sample Testing via Permutations} and Algorithm~\ref{Alg: Local Two-Sample Testing via Permutations}, respectively.

\begin{algorithm}[!htb]
	\caption{Global Two-Sample Testing via Permutations} 
	\label{Alg: Global Two-Sample Testing via Permutations} 
	{\small
		\begin{algorithmic}
			\REQUIRE samples $\{X_i, Y_i\}_{i=1}^n$, number of permutations $B$, significance level $\alpha$, a regression method. \\
			\STATE (1) Calculate the global test statistic $\widehat{\mT}_{global}$.
			\STATE (2) Randomly permute $\{Y_1,\ldots,Y_n\}$. Calculate the test statistic using the permuted data.
			\STATE (3) Repeat the previous step $B$ times to obtain $\big\{ \widehat{\mT}_{global}^{(1)},  \ldots, \widehat{\mT}_{global}^{(B)}\big\}$.
			\STATE (4) Approximate the permutation $p$-value by
			$$p = \frac{1}{B+1} \left( 1 + \sum_{b=1}^B I(\widehat{\mT}_{global}^{(b)} > \widehat{\mT}_{global}  ) \right).$$
			\STATE (5) Reject the null hypothesis when $p < \alpha$. Otherwise, accept the null hypothesis.
		\end{algorithmic}
	}
\end{algorithm}

\begin{algorithm}[!htb] 
	\caption{Local Two-Sample Testing via Permutations} 
	\label{Alg: Local Two-Sample Testing via Permutations}
	{\small
		\begin{algorithmic} 
			\REQUIRE samples $\{X_i, Y_i\}_{i=1}^n$, test points $\{x_j\}_{j=1}^k$, number of permutations $B$, 
			significance level $\alpha$, a multiple testing procedure, a regression method. \\
			\STATE (1) Calculate the test statistic $\widehat{\mT}_{local}(x_j)$ at the $k$ test points.
			\STATE (2) Randomly permute $\{Y_1,\ldots,Y_n\}$. Calculate the test statistic using the permuted data.
			\STATE (3) Repeat the previous step $B$ times to obtain $\{ \widehat{\mT}_{local}^{(1)}(x_j)\}_{j=1}^k, \ldots, \{\widehat{\mT}_{local}^{(B)}(x_j)\}_{j=1}^k$.
			\STATE (4) Approximate the permutation $p$-value at each test point $x_j$ by
			$$p_j = \frac{1}{B+1} \left( 1 + \sum_{b=1}^B I( \widehat{\mT}_{local}^{(b)}(x_j) > \widehat{\mT}_{local}(x_j)  ) \right).$$
			\STATE (5) Apply a multiple testing procedure for controlling the FWER or the FDR at $\alpha$ level.
			\STATE (6) Return the significant local test points.
		\end{algorithmic}
	}
\end{algorithm}

\subsection{Sampling Schemes} \label{Section: Sampling Schemes}
In the two-sample problem, there are two common sampling schemes for obtaining the paired data set $\{(X_i,Y_i)\}_{i=1}^n$, namely i) \emph{i.i.d.~sampling} and ii) \emph{separate sampling} defined as follows:
\begin{itemize}
	\item \textbf{i.i.d.~sampling.} Under i.i.d.~sampling, we observe $n$ pairs of i.i.d.~samples $\{(X_i,Y_i)\}_{i=1}^n$ from the joint distribution $\pi_1 [P_1 \times \delta_1] + \pi_0 [P_0 \times \delta_0]$. Here we note that $n$ is fixed in advance. Then $n_1 = \sum_{i=1}^n I(Y_i=1)$ and $n_0 = n - n_1$ are $\text{Binomial}(n,\pi_1)$ and $ \text{Binomial}(n,\pi_0)$, respectively. This setting is common in applications of supervised learning where the goal is to build a model that can successfully predict the class label $Y$ given the feature vector $X$ \citep[e.g.][]{friedman2009elements}. Our goal, on the other hand, is to test whether the two distributions $P_0$ and $P_1$ are the same or not by leveraging existing methods in the regression literature. \\[-.5em]
	\item \textbf{Separate sampling.} In the case of separate sampling, $n_0$ and $n_1$ are predetermined and they are not random. We then observe $n_0$ and $n_1$ independent sample points from $P_0$ and $P_1$ separately, which provides the data set $\{(X_i,Y_i)\}_{i=1}^n$ where $Y_i = 1$ if $X_i$ was drawn from $P_1$ and $Y_i=0$ otherwise. 
\end{itemize}

We can link the separate sampling to the i.i.d.~sampling scheme by randomly ordering the $(X_i, Y_i)$ pairs, so that the data points are exchangeable and for each $i \in \{1,\ldots,n\}$, the conditional distribution of $Y_i$ given $X_i = x$ is $m(x) = \pi_1 f_1(x) / \{ \pi_1 f_1(x)  + \pi_0 f_0(x)\}$ where the class probability is given by $\pi_1 = n_1/n$. Therefore, although the joint distributions of $\{(X_i,Y_i)\}_{i=1}^n$ are different under i.i.d.~and separate sampling schemes, they share the same regression function.



\begin{remark} \normalfont
	These two sampling schemes are also known as \emph{prospective sampling} and \emph{retrospective (or case-control) sampling}, respectively, and their relationships have been studied in different contexts. For example, it has been shown that the logistic slope estimates have similar behaviors under both sampling schemes \cite[see, e.g.][]{anderson1972separate,prentice1979logistic,wang1993robust,wang1999high,bunea2009dimension}. This result has been extended to general regression models by \cite{scott2001maximum}.
\end{remark}

\section{Global Two-Sample Tests via Regression}  \label{Section: Global Two-Sample Tests via Regression}
The choice of regression method in our framework will ultimately decide whether we achieve competitive statistical power. 
In Section~\ref{Section: Fisher's Linear Discriminant Analysis}, we illustrate the point that the global regression test can be optimal if we choose a suitable regression method. For this theoretical purpose, we focus on the regression test based on Fisher's LDA and show its optimality. In Section~\ref{Section: The MISE and Testing Error for Global Regression}, we turn our attention to more general regression settings and characterize the testing error of the global regression test in terms of the mean integrated square error (MISE) of the regression estimator.


\subsection{Fisher's Linear Discriminant Analysis} \label{Section: Fisher's Linear Discriminant Analysis}
In this section, we consider a simple scenario of two sample normal mean to highlight the difference between our approach and the classification accuracy approach. In particular, we prove that the regression test based on Fisher's LDA achieves the same local power as Hotelling's $T^2$ test. This result has significance given that i) Hotelling's test is optimal under the considered scenario and ii) the classification accuracy test based on Fisher's LDA is usually underpowered \citep{ramdas2016classification,rosenblatt2016better}. To facilitate comparison with the previous results, which are established under separate sampling, we also consider the case where $n_0$ and $n_1$ are predetermined throughout this subsection.


Suppose we observe $\{X_{i,0}\}_{i=1}^{n_0} \overset{i.i.d.}{\sim} N(\mu_0, \Sigma)$ and independently $\{X_{i,1}\}_{i=1}^{n_1} \overset{i.i.d.}{\sim} N(\mu_1, \Sigma)$. We denote the pooled samples by $\{X_i\}_{i=1}^n  =\{X_{i,0}\}_{i=1}^{n_0} \cup \{X_{i,1}\}_{i=1}^{n_1}$ where $n = n_0 + n_1$. The two-sample problem then becomes the problem of testing for mean differences as
\begin{align} \label{Eq: mean difference}
H_0: \mu_0 = \mu_1 \quad \text{versus} \quad H_1: \mu_0 \neq \mu_1. 
\end{align}
For this particular problem, Fisher's LDA is a natural choice for regression, assuming normality and equal class covariances. Let $\widehat{\mu}_i$ be the sample mean vector for each group, $\mathcal{S}$ be the covariance matrix of the combined samples, i.e. $\mathcal{S} = n^{-1} \sum_{i=1}^n (X_i - \widehat{\mu}) (X_i - \widehat{\mu})^\top$ where $\widehat{\mu} = n^{-1} \sum_{i=1}^n X_i $.  Then, by putting $\pi_1 = n_1/n$, the regression estimator based on Fisher's LDA is given by
\begin{align} \label{Fisher's LDA}
 \widehat{m}_{\text{LDA}}(x) =  \frac{ \pi_1 \exp \big\{ -\frac{1}{2}(x-\widehat{\mu}_1)^\top \mathcal{S}^{-1} (x- \widehat{\mu}_1) \big\} }{ \pi_0 \exp\big\{ -\frac{1}{2}(x- \widehat{\mu}_0)^\top \mathcal{S}^{-1} (x-\widehat{\mu}_0)  \big\} + \pi_1 \exp \big\{ -\frac{1}{2}(x- \widehat{\mu}_1)^\top \mathcal{S}^{-1} (x- \widehat{\mu}_1)  \big\}}. 
\end{align}
One of the most popular test statistics for testing (\ref{Eq: mean difference}) is Hotelling's $T^2$ statistic, which yields optimal power for the normal means problem \citep[see, e.g.][]{anderson2003introduction}. For the two-sample problem, Hotelling's $T^2$ statistic is defined by
\begin{align*}
T^2_{\text{Hotelling}} = \frac{n_0n_1}{n_0 + n_1} (\widehat{\mu}_0 - \widehat{\mu}_1)^\top \mathcal{S}_p^{-1} (\widehat{\mu}_0 - \widehat{\mu}_1),
\end{align*}
where $\mathcal{S}_p$ is the pooled covariance matrix, that is
\begin{align*}
\mathcal{S}_p = \frac{1}{n_0+n_1-2} \left( \sum_{i=1}^{n_0} (X_{i,0} - \widehat{\mu}_0)(X_{i,0} - \widehat{\mu}_0)^\top + \sum_{i=1}^{n_1} (X_{i,1} - \widehat{\mu}_1)(X_{i,1} - \widehat{\mu}_1)^\top \right).
\end{align*}
On the other hand, the regression test statistic based on Fisher's LDA is given by
\begin{align*}
\widehat{\mathcal{T}}_{\text{LDA}} =  \frac{1}{n}\sum_{i=1}^n \Big( \widehat{m}_{\text{LDA}}(X_i) - \pi_1 \Big)^2.
\end{align*}
The next theorem provides a connection between the seemingly unrelated $\widehat{\mathcal{T}}_{\text{LDA}}$ and $T^2_{\text{Hotelling}}$ statistics. Specifically, it shows that $n{\pi_0^{-1} \pi_1^{-1}} \widehat{\mathcal{T}}_{\text{LDA}}$ is asymptotically identical to Hotelling's $T^2$ statistic under the null. It is also worth pointing out that the theorem still holds without the normality assumption.

\begin{theorem} \label{Theorem: Asymptotic null distribution of Fisher's LDA}
	Let $\{X_{i,0}\}_{i=1}^{n_0}$ and $\{X_{i,1}\}_{i=1}^{n_1}$ be random samples under separate sampling from two multivariate distribution with the mean vectors $\mu_0$ and $\mu_1$, respectively, and the same covariance matrix $\Sigma$. Assume the pooled samples are mutually independent and the third moments of $X_{1,0}$ and $X_{1,1}$ are finite. Suppose that $\mathcal{S}_p$ and $\mathcal{S}$ satisfy $\mathcal{S}_p^{-1} = \Sigma^{-1}(1 + o_P(1))$ and $\mathcal{S}^{-1} = \Sigma^{-1}(1 + o_P(1))$. Then, under $H_0 : \mu_0 = \mu_1$, it holds that
	\begin{align} \label{Eq: LDA vs Hotelling}
	n\widehat{\mathcal{T}}_{\text{\emph{LDA}}} =n \pi_0^2 \pi_1^2 (\widehat{\mu}_0 - \widehat{\mu}_1)^\top \mathcal{S}_p^{-1} (\widehat{\mu}_0 - \widehat{\mu}_1) + o_P(1).
	\end{align}
	Therefore,
	\begin{align*}
	n{\pi_0^{-1} \pi_1^{-1}}  \widehat{\mathcal{T}}_{\text{\emph{LDA}}} = T_{\text{\emph{Hotelling}}}^2 + o_P(1) \overset{d}{\longrightarrow} \chi^2_D,
	\end{align*}
	where $\chi^2_D$ is the chi-squared distribution with $D$ degrees of freedom.
\end{theorem}

Let us now turn to the alternative hypothesis. To begin with, we consider a family of probability functions that satisfy the following smoothness condition.
\begin{definition}[Definition 12.2.1 of \citet{lehmann2006testing}]
	Let $\{P_\mu, \mu \in \Omega\}$ be a parametric model where $\Omega$ is an open subset of $\mathbb{R}^D$, and let $f_\mu(x)= dP_\mu(x) / d\nu(x)$ be the density function with respect to Lebesgue measure $\nu$. The family $\{P_\mu, \mu \in \Omega \}$ is quadratic mean differentiable (q.m.d.) at $\mu_0$ if there exists a vector of real-valued functions $\eta(\cdot, \mu_0) = \left( \eta_1(\cdot, \mu_0), \cdots, \eta_D(\cdot, \mu_0) \right)^\top$ such that 
	\begin{align}
	\int_{\mathbb{R}^D} \left[  \sqrt{f_{\mu_0 + h}(x)} - \sqrt{f_{\mu_0}(x)} - \langle \eta(x, \mu_0), h \rangle \right]^2  d\nu(x) = o(||h||_2^2)
	\end{align}
	as $||h||_2 \rightarrow 0$. 
\end{definition}
Such q.m.d.~families include fairly large parametric models such as exponential families in natural form. For our purpose, we focus on location q.m.d.~families, denoted by $\{\mP_\mu, \mu \in \Omega\}$. Specifically, $\mP_\mu$ is a member of  $\{\mP_\mu, \mu \in \Omega\}$ if its density satisfies $f_{\mu}(x) = f(x - \mu)$ for which $f(x)$ has zero mean and covariance matrix $\Sigma$. Next, for given $\mP_{\mu_0}$ and $\mP_{\mu_1}$ from $\{\mP_\mu, \mu \in \Omega\}$, let us consider the local alternative
\begin{align}  \label{Eq: localAlternative}
H_{1,n}: \mu_1- \mu_0 =  h / \sqrt{n} ,
\end{align}
where $h=(h_1,\ldots, h_D)^\top$. Then, under $H_{1,n}$, $\widehat{\mathcal{T}}_{\text{LDA}}$ has asymptotic behavior as follows.
\begin{theorem} \label{Theorem: Asymptotic distribution of LDA against local alternatives}
	Suppose under separate sampling that $\{X_{i,0}\}_{i=1}^{n_0} \overset{i.i.d.}{\sim} \mathbb{P}_{\mu_0}$ and independently $\{X_{i,1}\}_{i=1}^{n_1} \\ \overset{i.i.d.}{\sim} \mathbb{P}_{\mu_1}$ where $\mathbb{P}_{\mu_i}$ is a member of the location q.m.d.~family with the same covariance matrix $\Sigma$ and finite third moments. Suppose that $\mathcal{S}_p$ and $\mathcal{S}$ satisfy $\mathcal{S}_p^{-1} = \Sigma^{-1}(1 + o_P(1))$ and $\mathcal{S}^{-1} = \Sigma^{-1}(1 + o_P(1))$. Under the sequence of local alternatives given in (\ref{Eq: localAlternative}), we have
	\begin{align*}
	n{\pi_0^{-1} \pi_1^{-1}} \widehat{\mathcal{T}}_{\text{\emph{LDA}}} = T_{\text{\emph{Hotelling}}}^2 + o_P(1) \overset{d}{\longrightarrow} \chi^2_D(\lambda),
	\end{align*}
	where $\chi^2_D(\lambda)$ denotes a noncentral chi-square distribution with $D$ degrees of freedom and the noncentral parameter $$\lambda = \pi_0 \pi_1 h^\top \Sigma^{-1} h.$$
\end{theorem}

The results from Theorem \ref{Theorem: Asymptotic null distribution of Fisher's LDA} and Theorem \ref{Theorem: Asymptotic distribution of LDA against local alternatives} imply that our regression test based on $\widehat{\mathcal{T}}_{\text{LDA}}$ has the same asymptotic local power as Hotelling's $T^2$ test. As a result, the regression test based on $\widehat{\mathcal{T}}_{\text{LDA}}$ is asymptotically optimal against the local alternatives as Hotelling's $T^2$ test.

\begin{figure}[!t]
	\hfill
	\subfigure{\includegraphics[width=8.0cm]{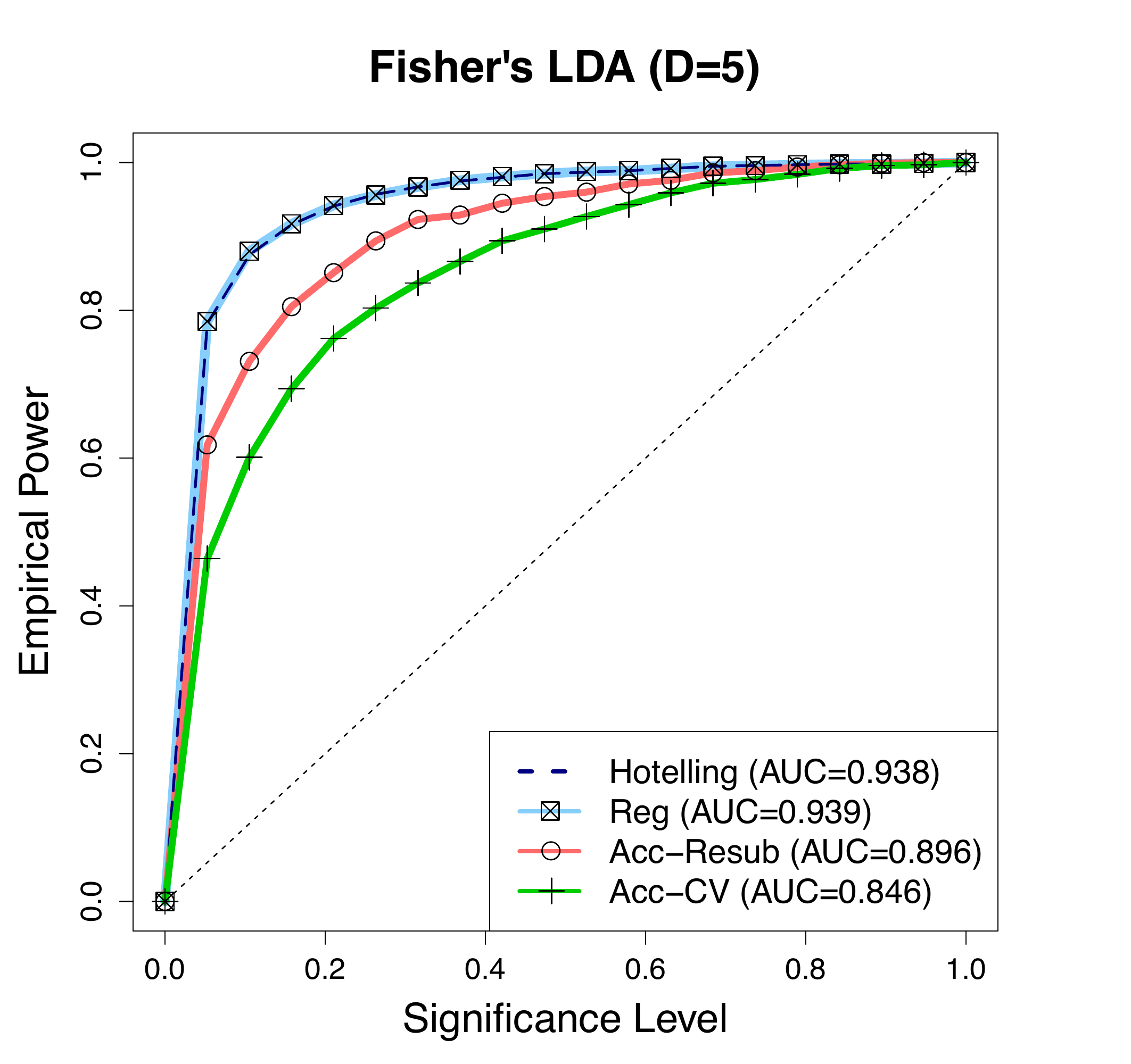}}
	\hfill
	\subfigure{\includegraphics[width=8.0cm]{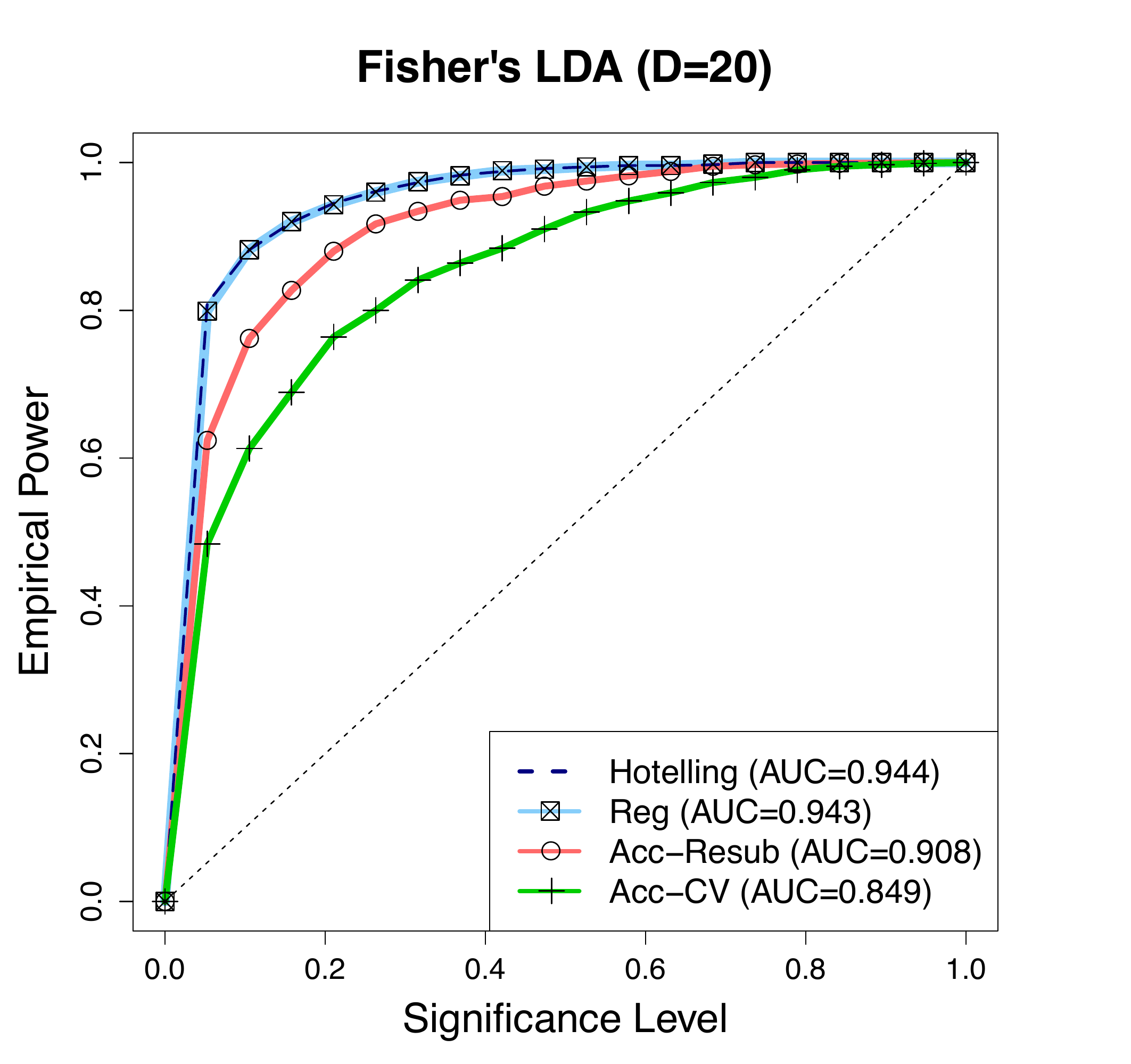}}
	\hfill
	\caption{Power comparisons between Hotelling's $T^2$~(Hotelling), $\widehat{\mathcal{T}}_{\text{LDA}}$~(Reg), the in-sample accuracy~(Acc-Resub), and the cross-validated accuracy~(Acc-CV) via Fisher's LDA.}  \label{Fig: LDA simulation}
\end{figure}

To illustrate the main point of this section, we compare the performance of $\widehat{\mathcal{T}}_{\text{LDA}}$ with Hotelling's $T^2$ test through Monte Carlo simulations. We randomly generate $n_0=n_1=100$ samples from $N((0,\ldots,0)^\top, I_D)$ and $ N((\mu,\ldots,\mu)^\top, I_D)$, respectively and set $\mu^2 = 0.05$ for $D=5$ and $\mu^2 = 0.01$ for $D=20$. We also consider two versions of the accuracy-based tests via Fisher's LDA: the in-sample (re-substitution) accuracy and the two-fold cross-validated accuracy. To calculate the cross-validated accuracy, we use the balanced sample splitting scheme in which the first part of data is used to train the LDA, and the second part is used to estimate the accuracy of the classifier \citep[see, Definition 1 and 2 of][for more details]{rosenblatt2016better}. To make a fair comparison, the critical values of the given tests were all decided by the permutation procedure. As shown in Figure \ref{Fig: LDA simulation}, the regression test based on $\widehat{\mathcal{T}}_{\text{LDA}}$ has comparable power to Hotelling's $T^2$ test that coincides with our theory. On the other hand, the accuracy tests have less power than Hotelling's $T^2$ test.

\subsection{The MISE and Testing Error for Global Regression} \label{Section: The MISE and Testing Error for Global Regression}
We now turn to more general regression settings and investigate general properties of the global regression test in both separate and i.i.d.~sampling cases. 
Let $\mathcal{M}$ be a certain class of regression $m(x): S \subseteq \mathbb{R}^D \mapsto [0,1]$ containing constant functions. Suppose that we have a regression estimator $\widehat{m}(x)$ that has the mean integrated square error as
\begin{align} \label{Eq: MSE}
\sup_{m \in \mathcal{M}} \mE \int_S \left( \widehat{m}(x) - m(x) \right)^2 dP_X(x) \leq C_0 \delta_n,
\end{align}
where $C_0$ is a positive constant and $\delta_n = o(1)$. In the case of i.i.d.~sampling, we further assume $\delta_n \geq n^{-1}$, which is typical for nonparametric regression estimators. Our main interest here is in employing the above MISE to characterize the testing error of the global regression test. Note that the plug-in global statistic in (\ref{Eq: Test Statistic}) is typically a biased estimator of the MISE and the bias differs from case to case. To simplify our analysis, we consider sample splitting where the half of data is used to estimate the regression function and the other is used to evaluate the empirical squared error. In detail, given samples $(X_1,Y_1), \ldots,  (X_{2n}, Y_{2n})$, the regression test statistic based on (random) sample splitting is defined by
\begin{align} \label{Eq: Global Test Statistic}
\widehat{\mathcal{T}}_{global}^\prime  = \frac{1}{n} \sum_{i=n+1}^{2n} \left( \widehat{m}(X_i) - \widehat{\pi}_1 \right)^2,
\end{align}
where $\widehat{m}(\cdot )$ and $\widehat{\pi}_1$ are calculated based on the first half of the data $\{(X_{1},Y_{1}), \ldots, (X_{n}, Y_{n}) \}$. In the case of separate sampling, we assume a random ordering in the entire data set and similarly split it into two parts but with the additional restriction that class probabilities are the same in both parts. Based on $\widehat{\mathcal{T}}_{global}^\prime$, we argue that for sufficiently large $C_1 > 0$ and $n$, the testing error of the global regression test can be arbitrarily small against the class of global alternatives given by
\begin{align*}
\mathcal{M}(C_1  \delta_n) = \Big\{ m \in \mathcal{M}: \int_S \left( m(x) - \pi_1 \right)^2 dP_X(x) \geq C_1 \delta_n \Big\}.
\end{align*}
Note that since $\pi_1$ is assumed to be fixed, the regression function $m(x)$ is completely determined by $f_0$ and $f_1$. Thus in the following theorem and hereafter, we use the notation $f_0,f_1 \in \mathcal{M}$ to represent $m(x) = \pi_1 f_1(x) / \{\pi_0 f_0(x) + \pi_1 f_1(x) \} \in \mathcal{M}$. Similarly, we write $f_0,f_1 \in \mathcal{M}_0$ to signify that $\pi_1 f_1(x) / \{\pi_0 f_0(x) + \pi_1 f_1(x) \}= \pi_1$ for all $x \in S$. With this notation in hand, we state the main theorem of this subsection. 
\begin{theorem} \label{Theorem: General Results of Global Regression Tests}
	Consider the case of i.i.d.~sampling or separate sampling. In each case, suppose that we have a regression estimator $\widehat{m}(\cdot)$ satisfying (\ref{Eq: MSE}). Let $t_\alpha$ be the upper $\alpha$ quantile of the permutation distribution of $\widehat{\mathcal{T}}_{global}^\prime $ based on $\widehat{m}(\cdot)$ where we permute the first half of labels. For fixed $\alpha \in (0,1)$ and $\beta \in (0,1-\alpha)$, we assume that there exists a positive constant $C_{0,\alpha}^\prime$ such that $\sup_{f_0,f_1 \in \mathcal{M}}\mP_{f_0,f_1} (t_\alpha < C_{0,\alpha}^\prime \delta_n) \geq 1 - \beta/2$. Then there exist positive constants $C_1$ and $N$ depending on $C_0,C_{0,\alpha}^\prime,\alpha,\beta$ such that
	\begin{align*}
	& \text{$\bullet$ Type I error: }  \sup_{f_0,f_1 \in \mathcal{M}_0} \mP_{f_0,f_1} \left( \widehat{\mathcal{T}}_{global}^\prime  > t_\alpha \right) \leq \alpha \quad  \text{and} \\
	& \text{$\bullet$ Type II error: } \sup_{n \geq N} \sup_{f_0,f_1 \in \mathcal{M}(C_1 \delta_n)} \mP_{f_0,f_1} \left( \widehat{\mathcal{T}}_{global}^\prime  \leq  t_\alpha \right) \leq \beta.
	\end{align*}
\end{theorem}



Theorem~\ref{Theorem: General Results of Global Regression Tests} uses the assumption that the permutation critical value of the regression test is uniformly bounded by $\delta_n$ (up to some constant factor) with high probability. We end this subsection with a class of regression estimators, which satisfy this assumption. Let us consider a class of regression estimators with the following representation:
\begin{align*}
\widehat{m}(x) = \sum_{i=1}^n w_i(x) Y_i,
\end{align*}
where $w_i(x) \geq 0$ and $\sum_{i=1}^n w_i(x) = 1$ for all $x$. In addition, we assume that $w_i(x)$ is a function of $\{X_1,\ldots, X_n \}$ but not $\{Y_1,\ldots,Y_n\}$. This class of estimators, often called linear smoothers, contains many popular regression methods such as k-nearest neighbor (kNN) regression, kernel regression and local polynomial regression. Focusing on linear smoothers, we provide the following corollary.

\begin{corollary} \label{Corollary: Global Regression Test}
	Consider the case of i.i.d.~sampling or separate sampling. In each case, let $\widehat{\mathcal{T}}_{global}^\prime $ be the global regression test statistic in (\ref{Eq: Global Test Statistic}) based on a linear smoother $\widehat{m}(\cdot)$ with the property in (\ref{Eq: MSE}). Let $t_\alpha$ be the upper $\alpha$ quantile of the permutation distribution of $\widehat{\mathcal{T}}_{global}^\prime $ where we permute the first half of labels. Then for fixed $\alpha \in (0,1)$ and $\beta \in (0,1-\alpha)$, there exist positive constants $C_1$ and $N$ depending on $C_0, \alpha, \beta$ such that
	\begin{align*}
	& \text{$\bullet$ Type I error: } \sup_{f_0,f_1 \in \mathcal{M}_0} \mP_{f_0,f_1} \left( \widehat{\mathcal{T}}_{global}^\prime  > t_\alpha \right) \leq \alpha \quad  \text{and} \\
	& \text{$\bullet$ Type II error: } \sup_{n \geq N}\sup_{f_0,f_1 \in \mathcal{M}(C_1 \delta_n)} \mP_{f_0,f_1} \left( \widehat{\mathcal{T}}_{global}^\prime  \leq t_\alpha \right) \leq \beta.
	\end{align*}
\end{corollary}

\subsection{Examples} \label{Section: Examples}
In the case of i.i.d.~sampling, the convergence rate $\delta_n$ of commonly used regression estimators have been well-established and these results can be directly used to study the testing error of the global regression test. We list several known results here. More examples can be found in \cite{gyorfi2002distribution}, \cite{tsybakov2009introduction} and \cite{devroye2013probabilistic}.
\begin{itemize}
	\item \textbf{kNN regression.}~~When $\mathcal{M}$ is a class of Lipschitz continuous functions, the convergence rate of kNN estimators satisfies $\delta_n =  n^{-2/(2+D)}$ \citep{gyorfi2002distribution}. This can be generalized to a H\"{o}lder space with smooth parameter $\beta$ in which the rate becomes $\delta_n = n^{-2\beta / (2\beta + D)}$ \citep{gyorfi2002distribution,ayano2012rates} for $0 < \beta \leq 1.5$. Furthermore, \cite{kpotufe2011k} shows that kNN estimators are adaptive to the intrinsic dimension $d \ll D$ under appropriate conditions. In this case, the convergence rate becomes much faster as $\delta_n = n^{-2/(2+d)} \ll n^{-2/(2+D)}$. \\[-.5em]
	\item \textbf{Kernel regression.}~~Kernel regression estimators also achieve the converge rate as $\delta_n =  n^{-2/(2+D)}$ for Lipschitz continuous functions and more generally as $\delta_n =  n^{-2\beta/(2\beta+D)}$ for a H\"{o}lder space with smooth parameter $0 < \beta \leq 1.5$ \citep{gyorfi2002distribution}. The adaptivity of kernel regression to the intrinsic dimension has been proved by \cite{kpotufe2013adaptivity}. Following their results, the convergence rate becomes $\delta_n =  n^{-2/(2+d)} \ll n^{-2/(2+D)}$ when there exists a low-dimensional structure in the data. \\[-.5em]
	\item \textbf{Local polynomial regression.}~~Let $\mathcal{M}$ be a Sobolev space with smoothness $\alpha$. Then local polynomial regression estimators has the convergence rate as $\delta_n = n^{-\alpha/(\alpha + d)}$ where $d$ is manifold dimension smaller than the original dimension $D$ \citep{bickel2007local}.  \\[-.5em]
	\item \textbf{Random forests regression.}~~For Lipschitz continuous functions, \cite{biau2012analysis} shows that the random forest estimator converges at rate $\delta_n = n^{-\frac{0.75}{s\log2  + 0.75}}$ where $s$ is the number of the relevant features. Hence, the convergence rate of the random forests becomes faster than $n^{-2/(2+D)}$ when $s \leq D/2$ under certain conditions. \cite{wager2015adaptive} use the guess-and-check forest algorithm to show that the convergence rate of the random forest is $\delta_n =  n^{-\log(\xi)/\log(2\xi)}$ where $\xi = 1/(1-3/4s)$.
\end{itemize}
To the best of our knowledge, there has been no detailed investigation of the regression estimation error under separate sampling. In this case, we cannot directly take advantage of existing results on regression. However, as the sample size becomes larger, the difference between i.i.d.~sampling and separate sampling becomes minor. Hence we expect that a reasonable regression estimator behaves similarly under both sampling schemes in large sample sizes, while a detailed analysis is necessary in future work. It is also worth noting that for certain regression methods, consistency results are not significantly affected by sampling scheme. For example, the consistency theory for $L_1$ penalized regression relies mainly on the assumption about a design matrix, which can be fulfilled under both sampling schemes \citep{van2008high,buhlmann2011statistics}. In such a case, the same convergence rate can be established under both sampling schemes.

\section{Local Two-Sample Tests via Regression} \label{Section: Local Two-Sample Tests via Regression}
The global two-sample test only answers the question whether two distributions are different, whereas in some applications, 
it would be more valuable to describe how these two distributions differ in a multivariate space. With this goal in mind, we now move on to the local two-sample problem and study general properties of the local regression test. 

\subsection{The MSE and Testing Error for Local Regression} \label{Section: The MSE and Testing Error for Local Regression}
We start by establishing similar results in Section~\ref{Section: The MISE and Testing Error for Global Regression} for local regression tests. Given a local point $x \in S$ of interest, suppose that a regression estimator has the mean square error such that
\begin{align} \label{Eq: Local MSE}
\sup_{m \in \mathcal{M}} \mE \left[ \left( \widehat{m}(x) - m(x) \right)^2 \right] \leq C_{0,x} \delta_{n,x},
\end{align}
where $C_{0,x}$ is a positive constant and $\delta_{n,x} = o(1)$. In addition, we assume $\delta_{n,x} \geq n^{-1}$ for i.i.d.~sampling. Then the next theorem shows that for sufficiently large $C_{1,x}$ and $n$, the local testing error based on the given regression estimator can be arbitrarily small against the class of local alternatives given by
\begin{align*}
\mathcal{M}(C_{1,x} \delta_{n,x}) = \Big\{ m \in \mathcal{M} : \left( m(x) - \pi_1 \right)^2 \geq C_{1,x} \delta_{n,x} \Big\}.
\end{align*}

\begin{theorem} \label{Theorem: Local Regression Test}
	Consider the case of i.i.d.~sampling or separate sampling. In each case, consider the local regression test statistic $\widehat{\mathcal{T}}_{local}(x)$ in (\ref{Eq: Test Statistic}) based on a linear smoother $\widehat{m}(x) = \sum_{i=1}^n w_i(x) Y_i$ with the property in (\ref{Eq: Local MSE}). Let $t_\alpha$ be the upper $\alpha$ quantile of the permutation distribution of $\widehat{\mathcal{T}}_{local}(x)$. Then for fixed $\alpha \in (0,1)$ and $\beta \in (0,1-\alpha)$, there exist positive constants $C_{1,x}$ and $N_x$ such that
	\begin{align*}
	& \text{$\bullet$ Type I error: } \sup_{f_0,f_1 \in \mathcal{M}_0}\mP_{f_0,f_1} \left( \widehat{\mathcal{T}}_{local}(x) > t_\alpha \right) \leq \alpha \quad  \text{and} \\
	& \text{$\bullet$ Type II error: } \sup_{n \geq N_x} \sup_{f_0,f_1 \in \mathcal{M}(C_{1,x} \delta_{n,x})} \mP_{f_0,f_1} \left( \widehat{\mathcal{T}}_{local}(x) \leq  t_\alpha \right) \leq \beta.
	\end{align*}
\end{theorem}

\begin{remark} \label{Remark: follow-up} \normalfont
	Although Theorem~\ref{Theorem: Local Regression Test} focuses on a linear smoother, the same conclusion holds for other regression methods as long as there exists a positive constant $C_{0,x,\alpha}$ such that the permutation critical value $t_\alpha$ is bounded above by $C_{0,x,\alpha} \delta_n$ with high probability (see Theorem~\ref{Theorem: General Results of Global Regression Tests} for a more formal statement).
\end{remark}

In order to keep things as simple and concrete as possible, we next focus on the Lipschitz class and analyze the optimality of the local regression tests from a minimax point of view. In the rest of this section (Section~\ref{Section: Minimax Optimality over the Lipschitz Class}--\ref{Section: Limiting Distribution of Local Permutation Test Statistics}), we concentrate on \emph{i.i.d.~sampling scheme} to take full advantage of known regression results. However, as we discussed in Section~\ref{Section: Examples}, similar results are expected to hold under separate sampling as well.

\subsection{Minimax Optimality over the Lipschitz Class} \label{Section: Minimax Optimality over the Lipschitz Class}
For a fixed constant $L>0$, let us denote the Lipschitz function class by
\begin{align*}
\mathcal{M}_{Lip} = \Big\{ m : |m(x) - m(y)| \leq L ||x - y||_2~ \text{for all $x,y \in S$} \Big\}.
\end{align*}
We also denote the collection of $\alpha$ level tests by $\Phi_{n,\alpha} = \{ \phi :  \sup_{f_0,f_1 \in \mathcal{M}_0}\mP_{f_0,f_1} (\phi = 1) \leq \alpha  \}$ and denote the class of Lipschitz local alternatives by
\begin{align} \label{Eq: Lipschitz Alternatives}
\mathcal{M}_{Lip}(\delta_{n,x}) = \Big\{  m \in \mathcal{M}_{Lip} : \left( m(x) - \pi_1 \right)^2 \geq  \delta_{n,x} \Big\}.
\end{align}
With this notation and fixed $\alpha \in (0,1)$ and $\beta \in (0,1-\alpha)$, the \emph{minimum separation} is defined by
\begin{align} \label{Eq: Minimum separation}
\delta_{n,x}^\star = \inf \Big\{ \delta_{n,x} : \inf_{\phi \in \Phi_{n,\alpha}} \sup_{f_0,f_1 \in \mathcal{M}_{Lip}(\delta_{n,x} )} \mP_{f_0,f_1} (\phi = 0)  \leq  \beta \Big\},
\end{align}
which is the smallest distance between $m(x)$ and $\pi_1$ such that the power becomes nontrivial. Then a test is called minimax rate optimal if it has power uniformly over $\mathcal{M}_{Lip}(\delta_{n,x})$ such that $\delta_{n,x} \asymp \delta_{n,x}^\star$.

In this section, we will investigate minimax rate optimality of local regression tests over the Lipschitz class under i.i.d.~sampling. First we formally state an upper bound for the local estimation error based on kNN and kernel regression in Example~\ref{Example: kNN MSE} and Example~\ref{Example: kernel MSE}, respectively. We then use these results to obtain the upper bound for the minimum separation in Corollary~\ref{Corollary: Local Testing Error of kNN and kernel}.

\begin{example}[kNN regression] \label{Example: kNN MSE}
	For a fixed point $x \in S$, list the data by $$(X_{1,n}(x),Y_{1,n}(x)), \ldots, (X_{n,n}(x),Y_{n,n}(x)),$$ where $X_{k,n}(x)$ is the $k$th nearest neighbor of $x$ and $Y_{k,n}(x)$ is its pair. Consider the kNN regression estimator
	\begin{align} \label{Eq: local kNN estimate}
	\widehat{m}_{kNN}(x) = \frac{1}{k_n} \sum_{i=1}^{k_n} Y_{(i,n)}(x),
	\end{align} 
	and assume that $\mP(X \in B_{x,\epsilon}) > \tau_x \epsilon^D$ where $B_{x,\epsilon}$ is a ball of radius $\epsilon >0$ centered at $x$ and $\tau_x > 0$. Then
	\begin{align*}
	\sup_{m \in \mathcal{M}_{Lip}} \mE \left[ \left( \widehat{m}_{kNN}(x) - m(x) \right)^2 \right] \leq \frac{1}{4k_n} +  L^2 \frac{2\Gamma(2/D)}{D \tau_x^{2/D}} \left(\frac{k_n}{n}\right)^{2/D},
	\end{align*}
	and for $k_n = n^{2/(2+D)}$,  we have
	\begin{align*}
	\sup_{m \in \mathcal{M}_{Lip}} \mE \left[ \left( \widehat{m}_{kNN}(x) - m(x) \right)^2 \right] \leq C_{0,x} n^{-\frac{2}{2+D}},
	\end{align*}
	where $C_{0,x} = 1/4 + L^2 \Gamma(2/D) D^{-1} \tau_x^{-2/D}$.
\end{example}

A similar result can be established for kernel regression estimators as follows.

\begin{example}[Kernel regression] \label{Example: kernel MSE}
	Given a kernel $K : S \mapsto [0,\infty)$, the kernel regression estimator at a fixed point $x$ is given by
	\begin{align} \label{Eq: kernel estimate}
	\widehat{m}_{ker}(x) = \frac{\sum_{i=1}^n Y_i K\left( \frac{x - X_i}{h_n} \right)}{\sum_{i=1}^n K\left(\frac{x-X_i}{h_n}\right)}.
	\end{align}
	Assume there exists $0 < r  < R$ and $0 < \lambda < 1$ such that 
	\begin{align*}
	\lambda I(x \in B_{0,r}) \leq K(x) \leq I(x \in B_{0,R})
	\end{align*}
	where $B_{0,\epsilon}$ is a ball of radius $\epsilon > 0$ centered at the origin. Further assume that $\mP(X \in B_{x,\epsilon}) > \tau_x \epsilon^D$ for some $\tau_x> 0$. Then
	\begin{align*}
	\sup_{m \in \mathcal{M}_{Lip}} \mE \left[ \left( \widehat{m}_{ker}(x) - m(x) \right)^2 \right] \leq \left( \frac{1+\lambda}{4\lambda^2 \tau_x r^D} +  \frac{2 e^{-1}}{\tau_x r^D}  \right)\frac{1}{n h_n^D}  +    L^2 R^2 h_n^2
	\end{align*}
	and for $h_n = n^{-2/(2+D)}$, 
	\begin{align*}
	\sup_{f_0,f_1 \in \mathcal{M}_{Lip}} \mE \left[ \left( \widehat{m}_{ker}(x) - m(x) \right)^2 \right] \leq C_{0,x} n^{-\frac{2}{2+D}}
	\end{align*}
	where $C_{0,x} = (1+\lambda)/(4\lambda^2 \tau_x r^D) + 2e^{-1}/(\tau_x r^D)  + L^2 R^2$.
\end{example}

\begin{remark} \label{Remark: Proof of Lemma kNN} \normalfont
	Example~\ref{Example: kNN MSE} and Example~\ref{Example: kernel MSE} are well-known and standard except that we keep track of the constant $C_{0,x}$ over the Lipschitz class. 
	Similar results exist in the literature but for slightly different settings. Hence, in Appendix~\ref{Section: Proofs}, we present detailed proofs for these two examples heavily building on \cite{gyorfi2002distribution}. The proofs will also be used to study the performance of the kNN and kernel local regression tests under the existence of intrinsic dimension in Proposition~\ref{Proposition: Local Testing Error of kNN and kernel for manifold data}.
\end{remark}

From the previous examples together with Theorem~\ref{Theorem: Local Regression Test}, we conclude that the minimum separation in (\ref{Eq: Minimum separation}) satisfies $\delta_{n,x}^\star \lesssim n^{-2/(2+D)}$.  We summarize this result in the following corollary.

\begin{corollary}[Upper bound] \label{Corollary: Local Testing Error of kNN and kernel}
	Let us denote the local kNN and kernel regression test statistics by
	\begin{align} \label{Eq: kNN and kernel local test statistic}
	\widehat{\mathcal{T}}_{kNN}(x) = (\widehat{m}_{kNN}(x) - \widehat{\pi}_1)^2, \quad \widehat{\mathcal{T}}_{ker}(x) = (\widehat{m}_{ker}(x) - \widehat{\pi}_1)^2,
	\end{align}
	and the upper $\alpha$ quantile of the permutation distribution of each statistic by $t_{\alpha, kNN}$ and $t_{\alpha, ker}$ respectively. Suppose the conditions in Example~\ref{Example: kNN MSE} holds with $k_n= n^{2/(D+2)}$. Then for fixed $\alpha \in (0,1)$ and $\beta \in (0,1-\alpha)$, there exist positive constants $C_{1,x}$ and $N_x$ such that
	\begin{align*}
	& \text{$\bullet$ Type I error: } \sup_{f_0,f_1 \in \mathcal{M}_0} \mP_{f_0,f_1} \left( \widehat{\mathcal{T}}_{kNN}(x) > t_{\alpha,kNN} \right) \leq \alpha \quad  \text{and} \\
	& \text{$\bullet$ Type II error: } \sup_{n \geq N_x} \sup_{f_0,f_1 \in \mathcal{M}_{Lip}(C_{1,x}n^{-2/( 2+ D)})} \mP_{f_0,f_1} \left( \widehat{\mathcal{T}}_{kNN}(x) \leq t_{\alpha,kNN} \right) \leq \beta.
	\end{align*}
	On the other hand, under the conditions in Example~\ref{Example: kernel MSE} with $h_n = n^{-2/(2+D)}$ and for fixed $\alpha \in (0,1)$ and $\beta \in (0,1-\alpha)$, there exist positive constants $C_{1,x}$ and $N_x$ such that
	\begin{align*}
	& \text{$\bullet$ Type I error: } \sup_{f_0,f_1 \in \mathcal{M}_0} \mP_{f_0,f_1} \left( \widehat{\mathcal{T}}_{ker}(x) >  t_{\alpha,ker} \right) \leq \alpha \quad  \text{and} \\
	& \text{$\bullet$ Type II error: } \sup_{n \geq N_x}\sup_{f_0,f_1 \in \mathcal{M}_{Lip}(C_{1,x} n^{-2/( 2+ D)})} \mP_{f_0,f_1} \left( \widehat{\mathcal{T}}_{ker}(x) \leq t_{\alpha,ker} \right) \leq \beta
	\end{align*}
	As a result, the minimum separation satisfies $\delta_{n,x}^\star \lesssim n^{-2/(2+D)}$. 
\end{corollary}

Next based on the standard technique to lower bound the testing error \citep[e.g.,][]{ingster1987minimax,baraud2002non}, we establish a lower bound for the minimum separation by $n^{-2/(2+D)} \lesssim \delta_{n,x}^\star$. This results matches with the upper bound in Corollary~\ref{Corollary: Local Testing Error of kNN and kernel}. Therefore, the tests in Corollary~\ref{Corollary: Local Testing Error of kNN and kernel} are minimax rate optimal and cannot be improved.

\begin{theorem}[Lower bound] \label{Theorem: Lower Bound}
	For any given $\alpha \in (0,1)$ and $\beta \in (1 - \alpha)$, there exists a constant $C_{1,x} >0$ such that
	\begin{align*}
	\inf_{\phi \in \Phi_{n,\alpha}} \sup_{f_0,f_1 \in \mathcal{M}_{Lip}(C_{1,x} n^{-2/(2+D)})} \mP_{f_0,f_1}(\phi = 0)  \geq ~  1- \alpha - \beta.
	\end{align*}
\end{theorem}

\vskip 0.5em

\begin{remark} \normalfont
	In the context of two-sample testing, it is sometimes more natural to make smoothness assumptions on densities $f_0$ and $f_1$ rather than on the regression function. Here we briefly discuss how to translate the smoothness condition on $f_0$ and $f_1$ into a condition on the regression function. Suppose that density functions $f_0$ and $f_1$ are uniformly bounded below by $c>0$ \citep[see, e.g.][for a similar assumption]{yang1999information}. Then some algebra shows that
		\begin{align*}
		|m(x) - m(y)| \leq \pi_0 c^{-1} |f_0(x) - f_0(y)| + \pi_1 c^{-1} |f_1(x) - f_1(y)|.
		\end{align*}
		In other words, if $f_0$ and $f_1$ are Lipschitz continuous (or more generally H\"{o}lder continuous), then the regression function is also Lipschitz continuous with a different Lipschitz constant. This means that our theoretical results will remain valid for the class of Lipschitz densities with the boundedness condition.
\end{remark}

\subsection{An Approach to Intrinsic Dimension} \label{Section: Adaptation to Intrinsic Dimension}
The previous results show that no test is uniformly powerful when the square distance between $m(x)$ and $\pi_1$ is order of $n^{-2/(2+D)}$; therefore it demonstrates the typical curse of dimensionality. Suppose that data $X \in S \subseteq \mathbb{R}^D$ has low intrinsic dimension $d$ which is smaller than the original dimension $D$ (e.g. manifold data). In this case, we would like to have a test whose performance only depends on intrinsic dimension and thus avoids the curse of dimensionality. For this purpose, we consider the homogeneous measure which captures local dimension of data.
\begin{definition}\citep[Definition 2 of ][]{kpotufe2011k} \label{Definition: Local Doubling Measure}
	Fix $x \in S \subseteq \mathbb{R}^D$, and $r>0$. Let $C>0$ and $1 \leq d < D$. The probability measure $\mP(\cdot)$ is $(C,d)$-homogeneous on $B_{x,r}$ if we have $\mP(X \in B_{x,r^\prime}) \leq C\epsilon^{-d} \mP(X \in B_{x,\epsilon r^\prime})$ for all $r^\prime \leq r$ and $0 < \epsilon < 1$.  
\end{definition}

Using Definition~\ref{Definition: Local Doubling Measure}, we reproduce Corollary~\ref{Corollary: Local Testing Error of kNN and kernel} and show that the performances of the local kNN and kernel regression tests depend on the intrinsic dimension instead of the original dimension.

\begin{proposition} \label{Proposition: Local Testing Error of kNN and kernel for manifold data}
	Consider the same notations as in Corollary~\ref{Corollary: Local Testing Error of kNN and kernel} and let $x \in S \subseteq \mathbb{R}^D$. Suppose the probability measure $\mP(\cdot)$ is $(C,d)$-homogeneous on $B_{x,r}$. Then for the kNN regression test with $k_n= n^{2/(2 + d)}$ and for any $\beta \in (0,1-\alpha)$, there exist positive constants $C_{1,x}$ and $N_x$ such that
	\begin{align*}
	& \text{$\bullet$ Type I error: } \sup_{f_0,f_1 \in \mathcal{M}_0} \mP_{f_0,f_1} \left( \widehat{\mathcal{T}}_{kNN}(x) > t_{\alpha,kNN} \right) \leq \alpha \quad  \text{and} \\
	& \text{$\bullet$ Type II error: } \sup_{n \geq N_x} \sup_{f_0,f_1 \in \mathcal{M}_{Lip}(C_{1,x} n^{-2/( 2+ d)})} \mP_{f_0,f_1} \left( \widehat{\mathcal{T}}_{kNN}(x) \leq t_{\alpha,kNN} \right) \leq \beta.
	\end{align*}
	On the other hand, for the kernel regression test with $h_n = n^{-2/(2+d)}$ and for any $\beta \in (0,1-\alpha)$, there exist positive constants $C_{1,x}$ and $N_x$ such that
	\begin{align*}
	& \text{$\bullet$ Type I error: } \sup_{f_0,f_1 \in \mathcal{M}_0} \mP_{f_0,f_1} \left( \widehat{\mathcal{T}}_{ker}(x) > t_{\alpha,ker} \right) \leq \alpha \quad  \text{and} \\
	& \text{$\bullet$ Type II error: } \sup_{n \geq N_x} \sup_{f_0,f_1 \in \mathcal{M}_{Lip}(C_{1,x} n^{-2/( 2+ d)})} \mP_{f_0,f_1} \left( \widehat{\mathcal{T}}_{ker}(x) \leq t_{\alpha,ker} \right) \leq \beta.
	\end{align*}
\end{proposition}

When the intrinsic dimension is unknown, one can employ a Bonferroni procedure to obtain the same results in Proposition~\ref{Proposition: Local Testing Error of kNN and kernel for manifold data}. To illustrate the idea, let $k_n(i) = n^{-2/(i+2)}$ for $i=1,\ldots,D$ and denote the resulting $kNN$ tests by $\phi_i(\alpha) = I(\mathcal{T}^{(i)}_{kNN}(x) > t_{\alpha,kNN}^{(i)})$ where $\mathcal{T}^{(i)}_{kNN}(x)$ and $t_{\alpha,kNN}^{(i)}$ are the kNN test statistic calculated with $k_n(i)$ and the corresponding $\alpha$ level permutation critical value, respectively. Then the final test is defined by $\phi_{max} = \max_{1 \leq i \leq D} \phi_i(\alpha/D)$. By using the union bound, it is easy to see that $\sup_{f_0,f_1 \in \mathcal{M}_0} \mP_{f_0,f_1} \left(  \phi_{max} = 1 \right) \leq \alpha$ and
\begin{align*}
\sup_{n \geq N_x} \sup_{f_0,f_1 \in \mathcal{M}_{Lip}(C_{1,x} n^{-2/( 2+ d)})} \mP_{f_0,f_1} \left( \phi_{max} = 0 \right) \leq \beta,
\end{align*}
for certain $C_{1,x}$ and $N_x$. This shows that the Bonferroni test does not lose any power in terms of separation rate and it adapts to the unknown intrinsic dimension. Despite this theoretical guarantee, the Bonferroni approach should be used with caution in practice. Indeed the Bonferroni test might be too conservative since it does not take into account the dependency structure among $\phi_1,\ldots,\phi_D$. 

\begin{remark} \normalfont
	For simplicity, we illustrate our idea on the Lipschitz class which only requires a mild smoothness assumption. Nevertheless our results in Section~\ref{Section: Minimax Optimality over the Lipschitz Class}--\ref{Section: Adaptation to Intrinsic Dimension} can be extended to a general function class such as H\"{o}lder class \citep[e.g.~Chapter 3.2 of][]{gyorfi2002distribution} in a similar way. Indeed, all we need is a uniform bound for the MSE (\ref{Eq: Local MSE}) over a general class, which can be found in the regression literature (See Section~\ref{Section: Examples}).
\end{remark}

\subsection{Limiting Distribution of Local Permutation Test Statistics} \label{Section: Limiting Distribution of Local Permutation Test Statistics}
When the sample size is large, calculating the permutation distribution is time-consuming. Hence it would be useful to investigate the limiting distribution of the permutation statistic. Based on the combinatorial central limit theorem~\citep[e.g.][]{bolthausen1984estimate}, we show that the permutation distribution of our local test statistic converges to the chi-square distribution with one degree of freedom as the sample size tends to infinity.

\begin{theorem} \label{Theorem: Limiting distribution of Local Regression Test}
	Consider the local regression test statistic $\widehat{\mathcal{T}}_{local}(x)$ in (\ref{Eq: Test Statistic}) based on a linear smoother $\widehat{m}(x) = \sum_{i=1}^n w_i(x) Y_i$. Suppose that
	\begin{align} \label{Eq: Condition for Permutation CLT}
	\frac{\max_{1 \leq i \leq n} |w_i(x) - 1/n|}{ \{ \sum_{i=1}^n (w_i(x) - 1/n)^2  \}^{1/2}} \convP 0
	\end{align}
	holds and let 
	\begin{align} \label{Eq: Sigma Expression}
	\sigma_n^2 = \frac{n}{n-1} \widehat{\pi}_1 (1 - \widehat{\pi}_1) \sum_{i=1}^n \left(w_i(x) - \frac{1}{n} \right)^2.
	\end{align}
	Further let $\eta = (\eta_1,\ldots,\eta_n)$ be a permutation of $\{1,\ldots,n\}$. Then the permutation distribution of the one-side local regression statistic converges to the standard normal distribution as
	\begin{align*}
	\sup_{t \in \mathbb{R}} \Big| \mP_\eta \left( \sigma_n^{-1} (\widehat{m}_\eta(x) - \widehat{\pi}_1) \leq t \Big| \mathcal{X}_n \right) - \mP\left( N(0,1) \leq t \right) \Big|  \convP 0.
	\end{align*}	
	Here $\mP_\eta(\cdot | \mathcal{X}_n)$ is the uniform probability measure over permutations conditioned on $(X_1,Y_1),\ldots,(X_n,Y_n)$ and  $\widehat{m}_\eta(x) = \sum_{i=1}^n w_i(x) Y_{\eta_i}$. Thereby, $\sigma_n^{-2} \widehat{\mathcal{T}}_{local}(x)$ converges to the chi-square distribution with one degree of freedom as
	\begin{align*}
	\sup_{t \in \mathbb{R}} \Big| \mP_\eta \left( \sigma_n^{-2} \widehat{\mathcal{T}}_{local}(x) \leq t \Big| \mathcal{X}_n \right) - \mP\left(  \chi^2_1 \leq t \right) \Big|  \convP 0.
	\end{align*}
\end{theorem}

We illustrate Theorem~\ref{Theorem: Limiting distribution of Local Regression Test} using kNN and kernel regression and show that both $\sigma_n^{-2} \widehat{\mathcal{T}}_{kNN}(x)$ and $\sigma_n^{-2} \widehat{\mathcal{T}}_{ker}(x)$ converge to the chi-square distribution with one degree of freedom under appropriate conditions. 

\begin{corollary}[kNN regression] \label{Corollary: kNN example}
	Consider the kNN estimator in (\ref{Eq: local kNN estimate}) with 
	\begin{align*}
	\sigma_n^2 = \widehat{\pi}_1 (1 - \widehat{\pi}_1) \frac{(n-1)(n-k)}{n^2k}.
	\end{align*}
	Then the permutation distribution of $\sigma_n^{-2} \widehat{\mathcal{T}}_{kNN}(x)$ converges to the chi-square distribution with one degree of freedom when $n, k \rightarrow \infty$ and $2k < n$.
\end{corollary}

\begin{corollary}[Kernel regression] \label{Corollary: kernel example}
	Consider the kernel regression estimator in (\ref{Eq: kernel estimate}) and assume that $\sup_t |K(t)| = \mathcal{K} < \infty$, $\int K^2(t) dt < \infty$ and $\int K_h(t)dx = 1$ where $K_h(t)= h^{-D} K(t/h)$. Denote the density function of $X$ by $f(\cdot)$. Assume that $0 < f(x) < \infty$ and $f(\cdot)$ is twice differentiable at $x$. Further assume that $nh^D \rightarrow \infty$ and $h \rightarrow 0$. Then the permutation distribution of $\sigma_n^{-2} \widehat{\mathcal{T}}_{ker}(x)$ converges to the chi-square distribution with one degree of freedom where $\sigma_n^{2}$ is given in (\ref{Eq: Sigma Expression}).
\end{corollary}

\section{Simulations} \label{Section: Simulations}

In this section, we carry out simulation studies for global and local two-sample tests to examine the empirical performance of the proposed methods. Throughout our simulations, we focus on the separate sampling scenarios under which other existing two-sample tests are usually investigated. We begin by comparing the regression test based on random forests \citep{breiman2001random} with other benchmark competitors in Section~\ref{Section: Random Forests Two-Sample Testing}. Next in Section~\ref{Section: A Comparison between Regression and Classification Accuracy Tests}, we illustrate by an example that the classification accuracy tests can fail due to their discrete nature while the corresponding regression tests perform well. We also provide simulation results for the local regression test in Section~\ref{Section: Toy Examples for Local Two-Sample Testing} to validate our approach.

\subsection{Random Forests Two-Sample Testing} \label{Section: Random Forests Two-Sample Testing}
Random forests have been proven to be a powerful tool for regression and classification problems in many application areas \citep[see e.g.,][]{hamza2005empirical, diaz2006gene, cutler2007random, chen2012random}. Despite the good performance of random forests in classification and regression problems, only a few works have applied these methods to statistical inference problems. To the best of our knowledge, only \cite{gagnon2016classification} and \cite{hediger2019use} use random forests for the two-sample problem. Now whereas \cite{gagnon2016classification} and \cite{hediger2019use} consider an accuracy test based on random forests, we propose a regression test based on random forests. The corresponding test statistic is given by
\begin{align} \label{Eq: Global RF statistic}
\widehat{\mT}_{RF} = \frac{1}{n} \sum_{i=1}^n \left( \widehat{m}_{RF}(X_i) - \widehat{\pi}_1 \right)^2,
\end{align}
where $\widehat{m}_{RF}$ is the regression estimator from the random forest algorithm.  For our simulation study, we implement both the RF accuracy and regression tests with the \texttt{randomForest} package (version 4.6-12) in \texttt{R} with default options for the parameters. 
We found in our simulation study that the in-sample classification accuracy of random forests is typically one even under the null case; therefore, the resulting test has no power against any alternative. For this reason, we instead estimate the classification accuracy from out-of-bag samples (which is a default option provided by the \texttt{randomForest} package). Throughout this section, we denote the accuracy test statistic based on random forests by $\widehat{\mA}_{RF}$.

\subsubsection{Simulation Setting}

Our simulations analyze two main settings. The first setting includes dense alternatives where the two distributions are different over a number of coordinates. The second setting, on the other hand, considers sparse alternatives where the two distributions differ in only a few coordinates. We carry out the simulations via the permutation procedure with $100$ random permutations, repeated 300 times for all test statistics. The significance level is controlled at $\alpha = 0.05$.

\vskip .8em

\noindent \textbf{Dense Alternatives.} For the dense alternatives, we draw random samples of size $n_0=n_1=20$ and dimension $D=5,20,50,100,150$ and $200$ from either multivariate normal distributions $N(\mu, \Sigma)$ or multivariate Cauchy distribution $\mathsf{C}(\mu, \Sigma)$ with different location $\mu$ and scale $\Sigma$ parameters.  We consider the following scenarios:

{
	\begin{itemize}
		\item 	\textbf{Dense Normal Location.} Test
		$N(0, I_D) ~ \text{versus} ~ N\left( \mu, I_D  \right)$, where $\mu = (0.2,0.2,\ldots, 0.2)^\top$.
		
		\vskip .5em
		
		\item \textbf{Dense Cauchy Location.} Test $\mathsf{C}(0, I_D) ~ \text{versus} ~  \mathsf{C} (\mu, I_D)$, where $\mu = (0.3,0.3,\ldots, 0.3)^\top$.
		
		\vskip .5em
		
		\item \textbf{Dense Normal Scale.} Test $N(0, I_D) ~ \text{versus} ~ N(0, J_D )$, where $J_D$ is a diagonal matrix whose diagonal elements are $(0.6,0.6,\ldots,0.6)^\top$.
		\vskip .5em
		
		\item \noindent \textbf{Dense Cauchy Scale.} Test  $\mathsf{C}(0, I_D) ~ \text{versus} ~ \mathsf{C} \left( 0, J_D  \right)$, where $J_D$ is a diagonal matrix whose diagonal elements are $(0.5,0.5,\ldots,0.5)^\top$.
	\end{itemize}
}

\vskip .8em

\noindent \textbf{Sparse Alternatives.} Similarly, we generate random samples with $n_0=n_1=20$ and $D=20,50,100,200,300$ and $400$ from either multivariate normal distributions or multivariate Cauchy distributions. We consider the following problems:

\vskip .8em

{
	\begin{itemize}
		\item \textbf{Sparse Normal Location.} Test
		$N(0, I_D) ~ \text{versus} ~ N (\mu, I_D)$, where $\mu = (2,0,\ldots, 0)^\top$.
		
		\vskip .5em
		
		\item \textbf{Sparse Cauchy Location.} Test  $\mathsf{C}(0, I_D) ~ \text{versus} ~  \mathsf{C} (\mu, I_D)$, where $\mu = (3,0,\ldots, 0)^\top$.
		
		\vskip .5em
		
		\item \textbf{Sparse Normal Scale.} Test  $N(0, I_D) ~ \text{versus} ~ N \left(0, J_D  \right)$, where $J_D$ is a diagonal matrix with diagonal elements  $(0.01,1,\ldots,1)^\top$.
		\vskip .5em
		
		\item \noindent \textbf{Sparse Cauchy Scale.} Test  $\mathsf{C}(0, I_D) ~ \text{versus} ~ \mathsf{C} \left( 0, J_D  \right)$, where $J_D$ is a diagonal matrix with diagonal elements  $(0.01,1,\ldots,1)^\top$.
	\end{itemize}
}

As a benchmark competitor, we consider the maximum mean discrepancy (MMD) test \citep{gretton2012kernel} based on
\begin{align} \label{Eq: MMD Test}
\text{MMD}_n^2 = -\frac{2}{n_0 n_1} \sum_{i,j=1}^{n_0,n_1} k(X_{i,0},X_{i,1})  + \frac{1}{n_0^2} \sum_{i,j=1}^{n_0} k(X_{i,0},X_{j,0}) + \frac{1}{n_1^2} \sum_{i,j=1}^{n_1} k(X_{i,0}, X_{j,0}),
\end{align}
where $k(x, y)$ is the Gaussian kernel with a bandwidth chosen by the median heuristic, i.e. $k(x,y) = \exp\left( -{||x-y||_2^2}/\sigma_{\text{median}}  \right)$ \citep[see,][for details]{gretton2012kernel}. We also consider the Energy test \citep{szekely2004testing,baringhaus2004new} based on
\begin{align} \label{Eq: Energy Test}
{\text{Energy}}_n  =  \frac{2}{n_0 n_1} \sum_{i,j=1}^{n_0,n_1} || X_{i,0} - X_{j,1} ||_2 - \frac{1}{n_0^2} \sum_{i,j=1}^{n_0} ||X_{i,0} - X_{j,0} ||_2 - \frac{1}{n_1^2} \sum_{i,j=1}^{n_1} ||X_{i,1} - X_{j,1} ||_2. 
\end{align}

\subsubsection{Simulation Results}

Tables \ref{Table: Dense Location}--\ref{Table: Sparse Scale} summarize our simulation results. We see from Table \ref{Table: Dense Location} and \ref{Table: Dense Scale} that MMD$_n$ and Energy$_n$ perform better than the regression test ($\widehat{\mT}_{RF}$) and the accuracy test ($\widehat{\mA}_{RF}$) against the dense normal location and scale alternatives. Indeed, MMD$_n$  and Energy$_n$ are known to be asymptotically optimal against the normal location alternative with the identity covariance matrix \citep{ramdas2015adaptivity}. However, they are both moment-based statistics, and hence sensitive to outliers. They are also based on the Euclidean metric. A major issue of the Euclidean and similar metrics is that they assign weights to the coordinates proportional to their scale without screening for irrelevant variables. Consequently, neither MMD$_n$  nor Energy$_n$ can properly deal with sparse alternatives, which explains their poor performance against the sparse location and scale alternatives. On the other hand, the base learner of the random forest algorithm is the decision tree. The usual splitting rule of decision trees is  invariant to absolute values \citep[see e.g., Chapter 9.2 of ][]{friedman2009elements}, which leads to  robustness against outliers. Random forests also have the ability to handle sparse alternatives by randomly selecting a few variables during the tree-growing process. By averaging each tree, random forests eventually put more weight on informative variables. In general, $\widehat{\mT}_{RF}$ and $\widehat{\mA}_{RF}$ are comparable to or more powerful than MMD$_n$  and Energy$_n$ under the sparse location and scale alternatives. Finally, we note from our simulations that the regression test $\widehat{\mT}_{RF}$ exhibits higher power than the accuracy test $\widehat{\mA}_{RF}$ for the dense as well as the sparse alternatives. 

\begin{table}[!h]
	\centering
	\renewcommand{\tabcolsep}{5.7pt}
	\caption{Power analysis against dense location alternatives at level $\alpha=0.05$}
	\label{Table: Dense Location}
	{\footnotesize
		\begin{tabular*}{\textwidth}{c | c l cccc | cccccc}
			\hline
			& \multicolumn{6}{c}{\emph{Normal Dense Location}}     & \multicolumn{6}{|c}{\emph{Cauchy Dense Location}}     \\ \hline
			$D$ & $5$   & $~~20$  & $50$  & $100$ & $150$ & $200$ & $5$   & $20$  & $50$  & $100$ & $150$ & $200$ \\ \hline 
			$\widehat{\mathcal{T}}_{\text{RF}}$  & 0.123 & 0.187 & 0.303 & 0.417 & 0.573 & 0.633 & \textbf{0.157} & \textbf{0.370} & \textbf{0.607} & \textbf{0.803} & \textbf{0.893} & \textbf{0.950} \\ 
			$\widehat{\mathcal{A}}_{\text{RF}}$ & 0.070 & 0.117 & 0.233 & 0.340 & 0.440 & 0.510 & 0.093 & 0.260 & 0.503 & 0.693 & 0.793 & 0.857 \\ 
			MMD$_n$     & \textbf{0.143} & \textbf{0.290} & \textbf{0.520} & \textbf{0.723} & \textbf{0.880} & \textbf{0.937} & 0.097 & 0.057 & 0.053 & 0.050 & 0.060 & 0.040 \\ 
			Energy$_n$  & \textbf{0.156} & \textbf{0.283} & \textbf{0.530} & \textbf{0.720} & \textbf{0.877} & \textbf{0.940} & 0.083 & 0.077 & 0.073 & 0.057 & 0.057 & 0.057 \\ \hline
		\end{tabular*}
	}
\end{table}

\begin{table}[!h]
	\centering
	\renewcommand{\tabcolsep}{5.7pt}
	\caption{Power analysis against dense scale alternatives at level $\alpha=0.05$}
	\label{Table: Dense Scale}
	{\footnotesize
		\begin{tabular*}{\textwidth}{c | c l cccc | cccccc}
			\hline
			& \multicolumn{6}{c}{\emph{Normal Dense Scale}}     & \multicolumn{6}{|c}{\emph{Cauchy Dense Scale}}     \\ \hline
			$D$& $5$   & $~~20$  & $50$  & $100$ & $150$ & $200$ & $5$   & $20$  & $50$  & $100$ & $150$ & $200$ \\ \hline 
			$\widehat{\mathcal{T}}_{\text{RF}}$  & 0.133 & 0.187 & 0.260 & 0.350 & 0.410 & 0.473 & 0.287 & \textbf{0.557} & \textbf{0.790} & \textbf{0.937} & \textbf{0.953} & \textbf{0.970} \\ 
			$\widehat{\mathcal{A}}_{\text{RF}}$ & 0.097 & 0.150 & 0.200 & 0.277 & 0.277 & 0.290 & 0.230 & 0.407 & 0.663 & 0.783 & 0.840 & 0.877 \\ 
			MMD$_n$     & \textbf{0.210} & \textbf{0.563} & \textbf{0.847} & \textbf{0.993} & \textbf{0.997} & \textbf{1.000} & \textbf{0.380} & 0.380 & 0.407 & 0.407 & 0.400 & 0.400 \\ 
			Energy$_n$  & 0.080 & 0.263 & 0.397 & 0.657 & 0.847 & 0.913 & 0.283 & 0.293 & 0.310 & 0.310 & 0.313 & 0.297 \\ \hline
		\end{tabular*}
	}
\end{table}

\begin{table}[!h]
	\centering
	\renewcommand{\tabcolsep}{5.7pt}
	\caption{Power analysis against sparse location alternatives at level $\alpha=0.05$}
	\label{Table: Sparse Location}
	{\footnotesize
		\begin{tabular*}{\textwidth}{c | c l cccc | cccccc}
			\hline
			& \multicolumn{6}{c}{\emph{Normal Sparse Location}}     & \multicolumn{6}{|c}{\emph{Cauchy Sparse Location}}     \\ \hline
			$D$& $20$   & $~~50$  & $100$  & $200$ & $300$ & $400$ & $20$   & $50$  & $100$  & $200$ & $300$ & $400$ \\ \hline 
			$\widehat{\mathcal{T}}_{\text{RF}}$  & 0.953 & 0.880 & \textbf{0.830} & \textbf{0.687} & \textbf{0.600} & \textbf{0.503} & \textbf{0.960} & \textbf{0.933} & \textbf{0.897} & \textbf{0.710} & \textbf{0.643} & \textbf{0.577} \\ 
			$\widehat{\mathcal{A}}_{\text{RF}}$ & 0.883 & 0.817 & 0.763 & 0.600 & 0.523 & 0.440 & 0.943 & 0.877 & 0.830 & 0.613 & 0.540 & 0.527 \\ 
			MMD$_n$      & \textbf{0.977} & \textbf{0.943} & 0.770 & 0.587 & 0.437 & 0.360 & 0.147 & 0.067 & 0.057 & 0.043 & 0.057 & 0.027 \\ 
			Energy$_n$  & \textbf{0.977} & \textbf{0.943} & 0.770 & 0.587 & 0.440 & 0.367 & 0.157 & 0.083 & 0.043 & 0.037 & 0.050 & 0.040 \\ \hline
		\end{tabular*}
	}
\end{table}

\begin{table}[!h]
	\centering
	\renewcommand{\tabcolsep}{5.7pt}
	\caption{Power analysis against sparse scale alternatives at level $\alpha=0.05$}
	\label{Table: Sparse Scale}
	{\footnotesize
		\begin{tabular*}{\textwidth}{c | c l cccc | cccccc}
			\hline
			& \multicolumn{6}{c}{\emph{Normal Sparse Scale}}     & \multicolumn{6}{|c}{\emph{Cauchy Sparse Scale}}     \\ \hline
			$D$& $20$   & $~~50$  & $100$  & $200$ & $300$ & $400$ & $20$   & $50$  & $100$  & $200$ & $300$ & $400$ \\ \hline 
			$\widehat{\mathcal{T}}_{\text{RF}}$  &\textbf{0.630} & \textbf{0.333} & \textbf{0.287} & \textbf{0.167} & \textbf{0.167} & \textbf{0.133} & \textbf{0.830} & \textbf{0.550} & \textbf{0.390} & \textbf{0.257} & \textbf{0.197} & \textbf{0.170} \\ 
			$\widehat{\mathcal{A}}_{\text{RF}}$ & 0.603 & 0.297 & 0.220 & 0.130 & 0.120 & 0.087 & 0.743 & 0.467 & 0.287 & 0.207 & 0.170 & 0.150 \\ 
			MMD$_n$      & 0.043 & 0.057 & 0.043 & 0.053 & 0.060 & 0.063 & 0.067 & 0.033 & 0.040 & 0.057 & 0.063 & 0.043 \\ 
			Energy$_n$  & 0.037 & 0.050 & 0.043 & 0.050 & 0.060 & 0.063 & 0.047 & 0.047 & 0.040 & 0.057 & 0.053 & 0.037 \\ \hline
		\end{tabular*}
	}
\end{table}

\subsection{A Comparison between Regression and Classification Accuracy Tests} \label{Section: A Comparison between Regression and Classification Accuracy Tests}

As mentioned earlier, many classifiers are typically estimated by dichotomizing regression estimators. Depending on the alternative, this dichotomization can result in a less powerful accuracy test than the corresponding regression test. We specifically demonstrate this point by considering two commonly used nonparametric regression methods; namely, $k$-nearest neighbors regression and kernel regression.

\subsubsection{Simulation Setting}
Recall the kNN estimator and the kernel regression estimator in (\ref{Eq: local kNN estimate}) and (\ref{Eq: kernel estimate}), respectively. Using these estimators, the global regression test statistics are given by
\begin{align*}
\widehat{\mT}_{kNN} = \frac{1}{n} \sum_{i=1}^n \Big( \widehat{m}_{kNN}(X_i) - \widehat{\pi}_1  \Big)^2 \quad \text{and} \quad \widehat{\mT}_{ker} = \frac{1}{n} \sum_{i=1}^n \Big( \widehat{m}_{ker}(X_i) - \widehat{\pi}_1  \Big)^2.
\end{align*}
Here we use the Euclidean distance to measure the pairwise distance between observations for kNN. On the other hand, we consider the Gaussian kernel with a diagonal bandwidth matrix with identical components $h$ for kernel regression. The corresponding accuracy test statistics are
\begin{align*}
& \widehat{\mA}_{kNN} = \frac{1}{n} \sum_{i=1}^n I \Big(I ( \widehat{m}_{kNN}(X_i) > 1/2) = Y_i \Big) \quad \text{and} \quad  \widehat{\mA}_{ker} = \frac{1}{n} \sum_{i=1}^n I \Big(I ( \widehat{m}_{ker}(X_i) > 1/2) = Y_i \Big),
\end{align*}
respectively. For all tests, we reject the null hypothesis when the test statistic is larger than a permutation critical value. 

For the simulation study, we let $\{X_{1,0},\ldots,X_{n_0,0}\} \overset{i.i.d.}{\sim} N(\mu_0, \sigma_0^2 \times I_D)$ and $\{ X_{1,1},\ldots,X_{1,n_1}\} \overset{i.i.d.}{\sim} N(\mu_1, \sigma_1^2 \times I_D)$ where $\mu_0 = (0,\ldots, 0)^\top$, $\mu_1 = (0.2,\ldots, 0.2)^\top$, $\sigma_0^2 =1$, and $\sigma_1^2 = 1.2$. Hence, there exist differences in both the location and scale parameters. We choose the sample sizes $n_0=n_1=50$ and change the dimension from $D=5$ to $D=75$ by steps of 10. To compare the performance, we carry out the permutation test with $200$ permutations, and the simulations are repeated 1,000 times to estimate the power of the test. We provide results for a range of different values of the tuning parameters: $k=5,15,25$ for the $k$-NN regression, and $h=5,15,25$ for the kernel regression. 

\begin{figure}[t!]
	\begin{center}
		\includegraphics[width=\textwidth]{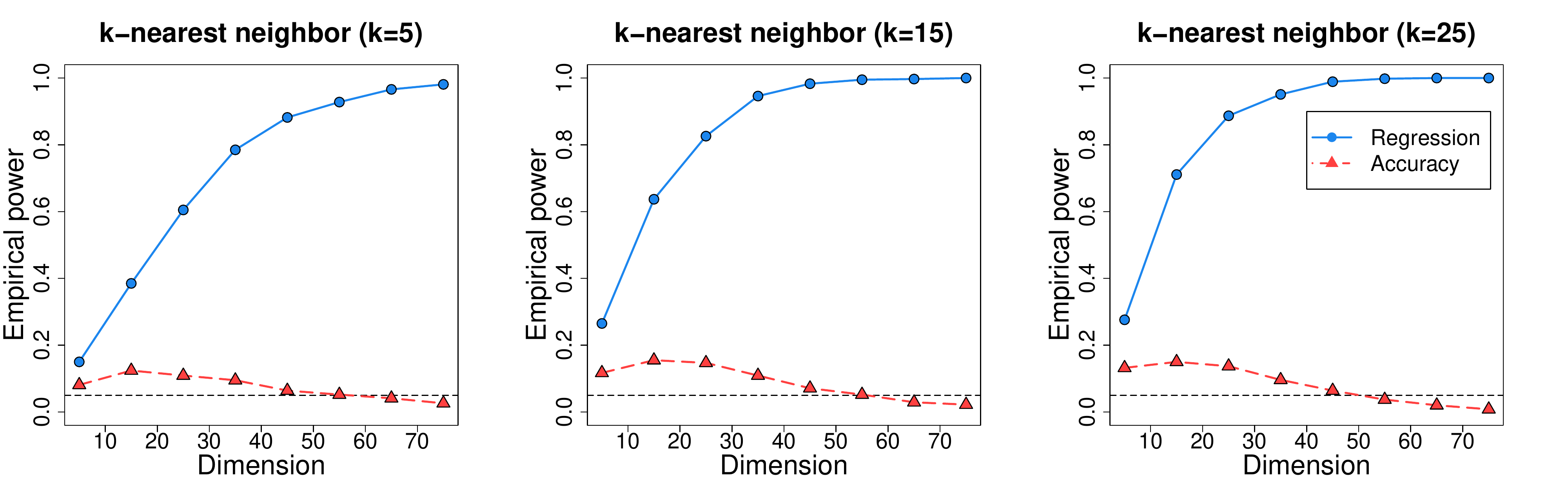}
		\caption{Power comparison between the regression test and the classification accuracy test via $k$-NN regression at level $\alpha=0.05$ for the toy example in Section~\ref{Section: A Comparison between Regression and Classification Accuracy Tests}.}\label{Fig: kNN Reg}
	\end{center}
	\begin{center}
		\includegraphics[width=\textwidth]{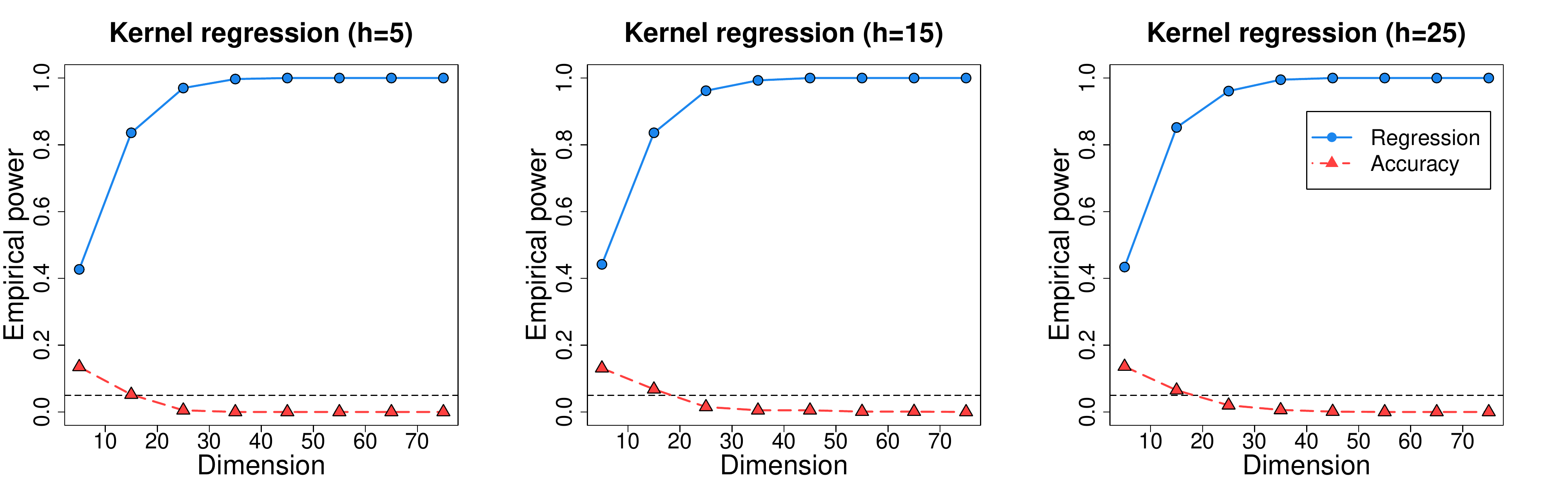}
		\caption{Power comparison between the regression test and the classification accuracy test via kernel regression at level $\alpha=0.05$ for the toy example in Section~\ref{Section: A Comparison between Regression and Classification Accuracy Tests}.} \label{Fig: Kernel Reg}
	\end{center}
\end{figure}

\subsubsection{Simulation Results}
Simulation results are presented in Figure~\ref{Fig: kNN Reg} and \ref{Fig: Kernel Reg}. From the results, it is seen that the regression tests consistently outperform the corresponding classification accuracy tests under the given scenario. The power of the accuracy tests even decreases with dimension, whereas the power of the regression tests steadily increases with dimension. The increase in power with dimension is desirable under this scenario because each coordinate presents evidence towards the alternative. The counter-intuitive result for the accuracy tests is due to the fact that the tests employ a dichotomized regression estimator. To explain it more clearly, we borrow some results from \cite{mondal2015high}. First, it can be shown by the weak law of large numbers that
\begin{align*}
& \text{1)} \ D^{-1/2} || X_{i,0} - X_{j,0} ||_2 \ \convP \ \sigma_0 \sqrt{2} \quad \text{for } 1 \leq i<j \leq n_0, \\[.5em]
& \text{2)} \ D^{-1/2} || X_{i,1} - X_{j,1} ||_2 \ \convP \ \sigma_1 \sqrt{2} \quad \text{for } 1 \leq i<j \leq n_1, \\[.5em]
& \text{3)} \ D^{-1/2} || X_{i,0} - X_{j,1} ||_2 \ \convP \ \sqrt{\sigma_0^2 + \sigma_1^2 + (\mu_0 - \mu_1)^2}
\end{align*}
for $1 \leq i \leq n_0, \ 1 \leq j \leq n_1$, as $D \rightarrow \infty$ while $n_0$ and $n_1$ are fixed. For the given example, we have $\sigma_0 \sqrt{2} < \sqrt{\sigma_0^2 + \sigma_0^2 + (\mu_0 - \mu_1)^2} < \sigma_1 \sqrt{2}$, which implies that every instance is closer to an instance from the class $Y=0$ than to other instances from the class $Y=1$. In other words, the nearest neighbors of any observation are most likely to be from the class $Y=0$. Note that both $k$-NN and kernel regression, explicitly or implicitly, use the Euclidean distance to calculate the proximity between two instances. Therefore, we observe with high probability that $\widehat{m}_{kNN}(X_i)$ and $\widehat{m}_{KerR}(X_i)$ are estimated as less than half and the dichotomized classifiers become
\begin{align*}
I\left(\widehat{m}_{kNN}(X_i) > 1/2 \right) = \mathcal{I}\left(\widehat{m}_{KerR}(X_i) > 1/2 \right) = 0, \quad \text{for all } i =1,\ldots,n.
\end{align*}
Due to this dichotomization, $\widehat{\mA}_{kNN}$ and $\widehat{\mA}_{KerR}$ converge to the empirical class probability $n_0/n$ under the alternative, resulting in poor power performance. On the other hand, the regression tests based on $\widehat{\mT}_{kNN}$ and $\widehat{\mT}_{ker}$ can be powerful as long as $\widehat{m}_{kNN}(x)$ and $\widehat{m}_{ker}(x)$ significantly deviate from the class probability. This is indeed the case under the considered scenario and thus explains why the regression tests outperform the corresponding classification tests.

\subsection{Toy Examples for Local Two-Sample Testing} \label{Section: Toy Examples for Local Two-Sample Testing}
Contrary to classification accuracy, our regression approach naturally leads to a local two-sample testing framework that provides further information on pointwise differences between two populations. We consider two toy examples to demonstrate the empirical performance of the local regression test. For the simulation study, we focus on the local kNN regression statistic in (\ref{Eq: kNN and kernel local test statistic}) with $k_n = n^{2/(2+D)}$ for the normal mixture example and $k_n = n^{2/(2+d)}$ for the manifold example. For both examples, we control the family-wise error rate (FWER) at $\alpha = 0.05$ via the Hochberg step up procedure \citep{hochberg1988sharper}.

\subsubsection{Normal mixture example} \label{Section: Normal mixture example}
\begin{figure}[!t]
	\begin{center}
		\subfigure{\includegraphics[width=6.0cm]{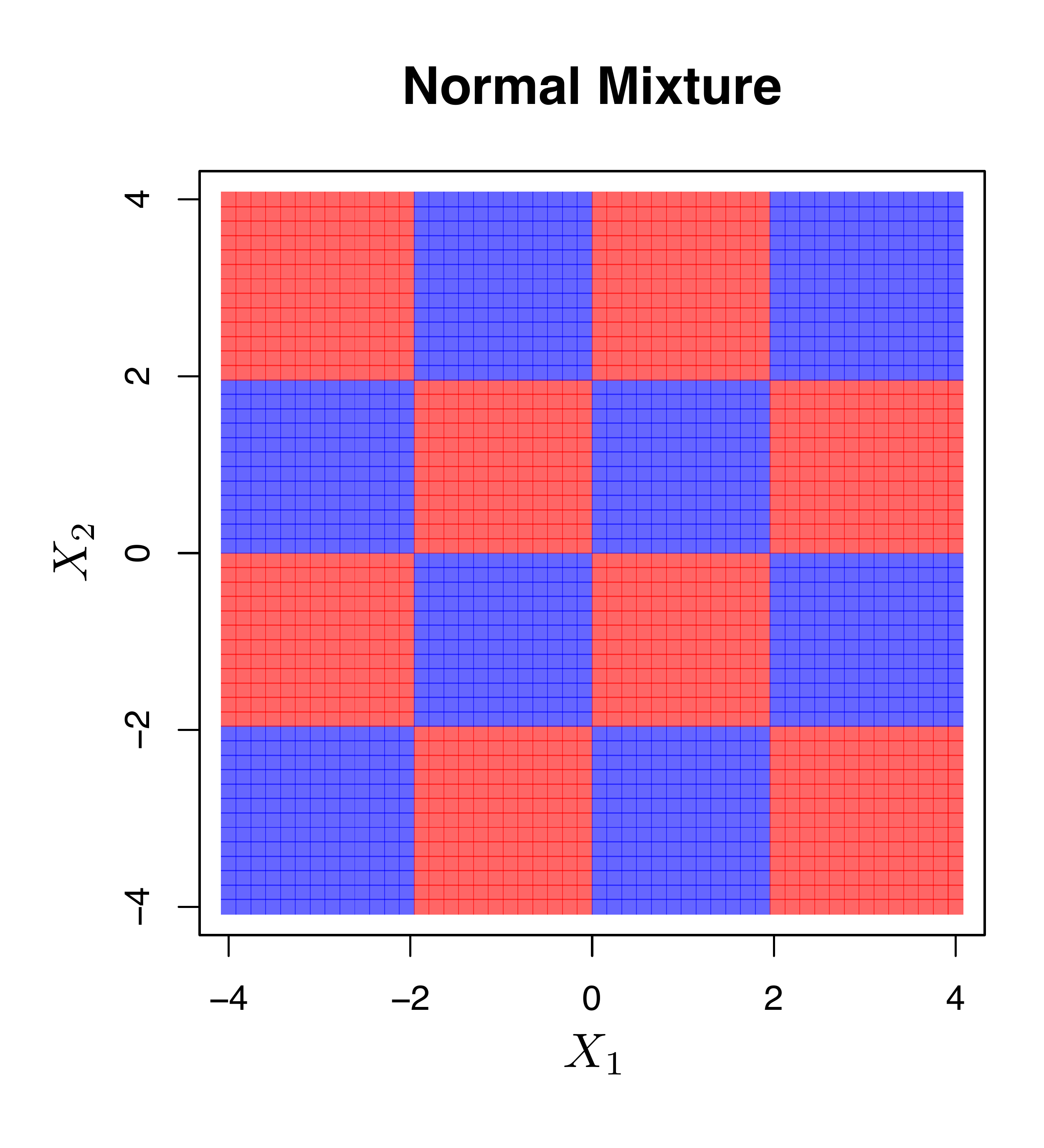}}
		\hskip 5em
		\subfigure{\includegraphics[width=6.0cm]{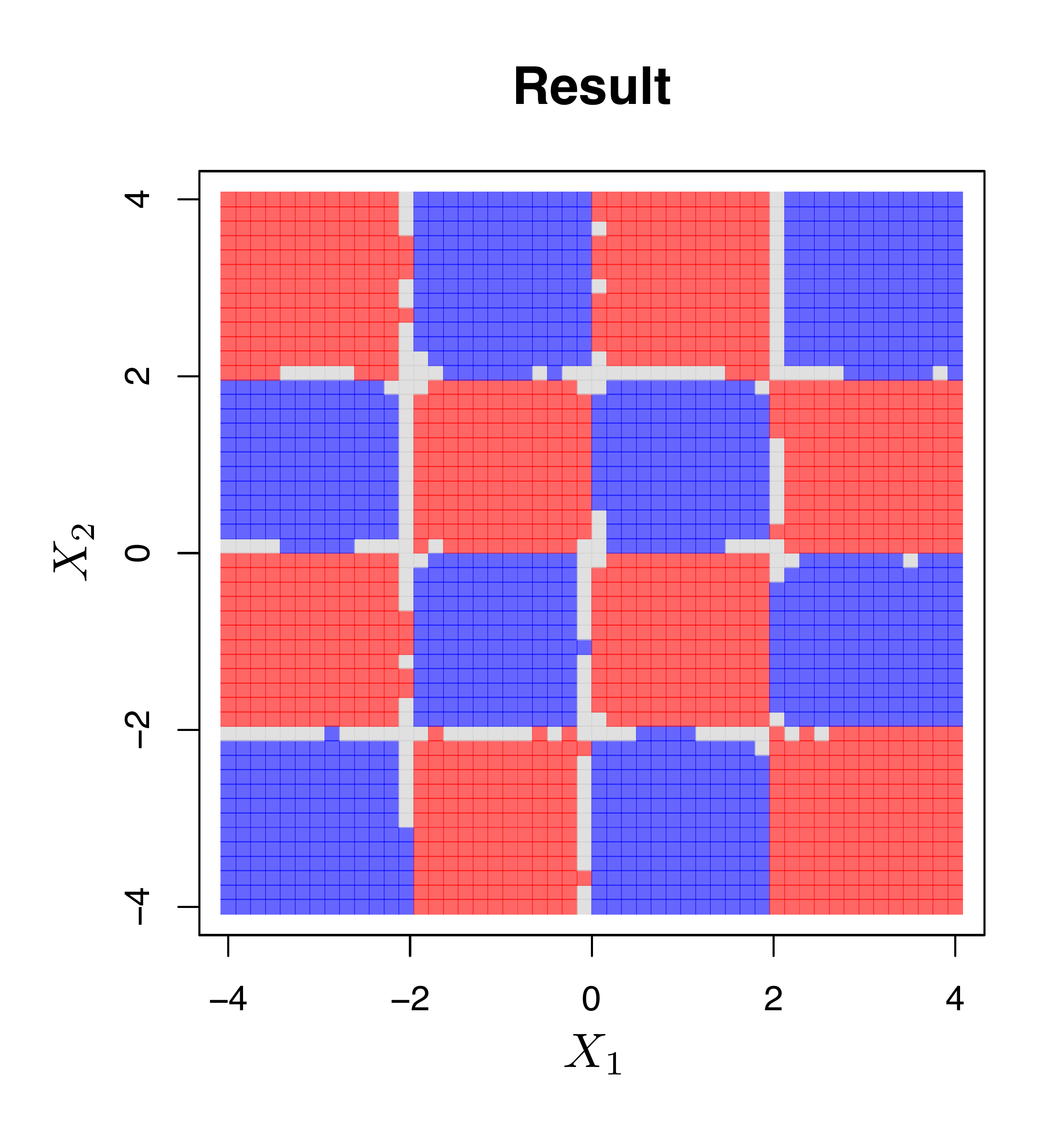}}
		\caption{Significant local regions for the normal mixture example. The left is the underlying true model and the right is the result of the local two-sample test. The difference regions are colored as follows --- (a) red: $f_1(x,y) > f_0(x,y)$, (b) blue: $f_1(x,y) < f_0(x,y)$ and (c) gray: insignificant.} \label{Figure: Normal mixture example}
	\end{center}
\end{figure}

In the first example, we consider two normal mixtures in $\mathbb{R}^2$:
\begin{align*}
f_0(x,y) = \frac{1}{8}\sum_{i=1}^8 \phi_i(x,y) \quad \text{and} \quad f_1(x,y) = \frac{1}{8}\sum_{i=1}^8 \phi_i^\prime(x,y),
\end{align*}
where $\phi_i$ is the bivariate normal density function with means $\mu_1 = (-3,-3)$, $\mu_2 = (-3,1)$, $\mu_3 = (-1,-1)$, $\mu_4 = (-1,3)$, $\mu_5 = (1,-3)$, $\mu_6 = (1,1)$, $\mu_7 = (3,-1)$, $\mu_8 = (3,3)$ and covariance matrix $\Sigma = 0.3^2 \times I_2$.  $\phi_i^\prime$ is similarly defined with means $\mu_1^\prime = (-3,-1)$, $\mu_2^\prime = (-3,3)$, $\mu_3^\prime = (-1,-3)$, $\mu_4^\prime = (-1,1)$, $\mu_5^\prime = (1,-1)$, $\mu_6^\prime = (1,3)$, $\mu_7^\prime = (3,-3)$, $\mu_8^\prime = (3,1)$ and the same covariance matrix. We generated $n_0 = n_1 = 2000$ samples from $f_0$ and $f_1$ and implemented Algorithm~\ref{Alg: Local Two-Sample Testing via Permutations} to capture local significant points. The local tests were performed at a fixed uniform grid of $50 \times 50$ points over $(x,y) \in [-4,4] \times [-4,4]$ and the result is presented in Figure~\ref{Figure: Normal mixture example}.

\subsubsection{Manifold data example} \label{Section: Manifold data example}
\begin{figure}[!t]
	\begin{center}
		\subfigure{\includegraphics[width=6.0cm]{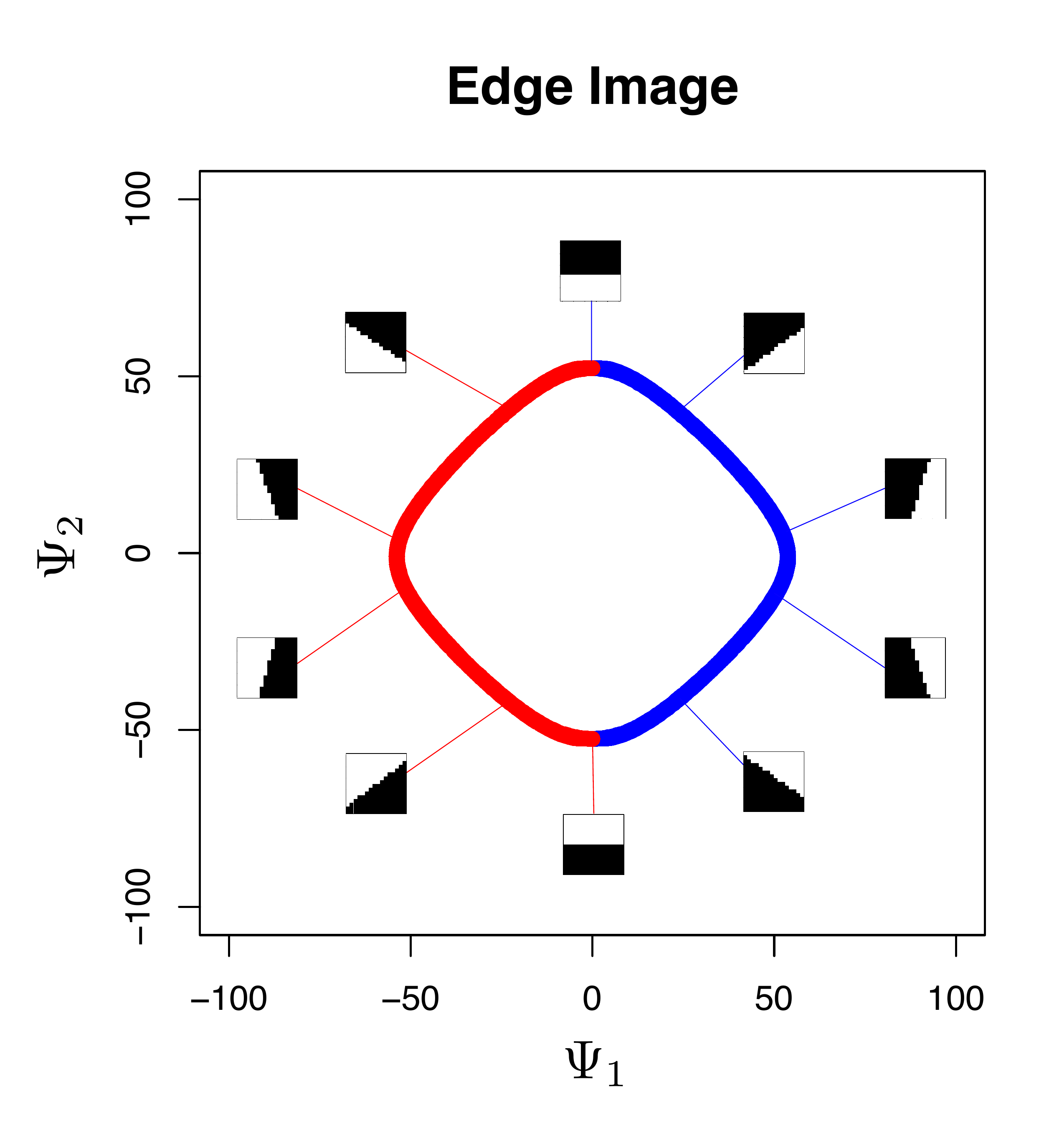}}
		\hskip 5em
		\subfigure{\includegraphics[width=6.0cm]{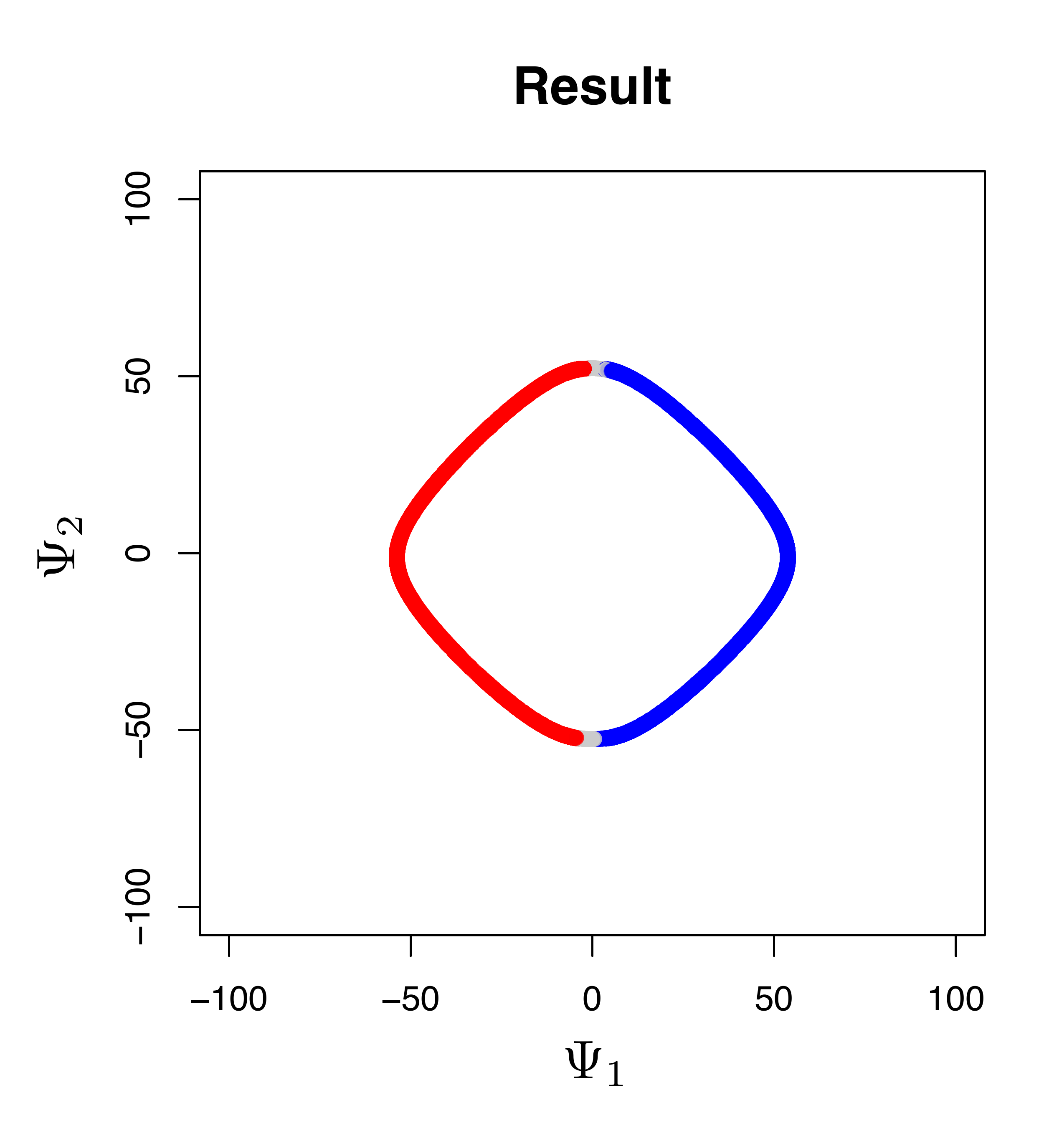}}
		\caption{Significant local regions for the manifold data example. The left is the underlying true model and the right is the result of the local two-sample test. The difference regions are colored as follows --- (a) red: $f_1(x_1,\ldots,x_{256}) > f_0(x_1,\ldots,x_{256})$, (b) blue: $f_1(x_1,\ldots,x_{256}) < f_0(x_1,\ldots,x_{256})$ and (c) gray: insignificant. Here $\Psi_1$ and $\Psi_2$ denote the the first two coordinates of the diffusion map.} \label{Figure: Edge image example}
	\end{center}
\end{figure}

In the second example, we create high-dimensional data with a low-dimensional manifold structure by generating edge images of size $16 \times 16$. Let $x,y$ be integers on evenly spaced points between $-30$ and $30$ that are $2$ units apart. Hence the size of the domain of $(x,y)$ becomes $16 \times 16$. Given two underlying parameters $\theta \in [-\pi, \pi]$ and $\rho \in [-5,5]$, an edge image is defined by
\begin{align*}
\mathcal{I}(x,y) = I\left( x \cdot \cos(\theta) + y \cdot \sin(\theta) - \rho > 0 \right).
\end{align*}
For the simulation, we draw $n_0=n_1=100$ samples from 
\begin{align*}
& (\theta_0, \rho_0) \sim \frac{1}{10}\text{Unif}([0, \pi] \times [0,5]) +  \frac{9}{10}\text{Unif}([-\pi, 0] \times [-5,0])\quad \text{and} \\[.5em]
& (\theta_1, \rho_1) \sim \frac{9}{10}\text{Unif}([0, \pi] \times [0,5]) +  \frac{1}{10}\text{Unif}([-\pi, 0] \times [-5,0]), 
\end{align*}
and generate corresponding edge images. As a result, there are two sets of the edge images supported on $\mathbb{R}^{256}$. Using these image samples, we implemented Algorithm~\ref{Alg: Local Two-Sample Testing via Permutations} to detect local significant points. The local tests were performed at fixed images whose parameters are defined on a uniform grid of $200 \times 200$ points over $(\theta, \rho) \in [-\pi,\pi] \times [-5,5]$. For visualization purpose, we projected the testing points into the two-dimensional diffusion space (see Appendix~\ref{Sec: Diffusion Map} for details) and the final result is provided in Figure~\ref{Figure: Edge image example}.

For both examples, the kNN local regression test performs reasonably well and detects most of the local differences between two distributions.

\section{Application to Astronomy Data} \label{Section: Application to Astronomy Data}

Continuing our discussion from Section~\ref{Section: Motivating Example}, we apply our two-sample framework to galaxies in the COSMOS, EGS, GOODS-North and UDS fields observed by the Hubble Space Telescope (HST) as part of the CANDELS program.\footnote{\href{http://candels.ucolick.org}{http://candels.ucolick.org}} For the analysis, we compute seven morphological statistics that summarize galaxy images nonparametrically: \emph{M, I, D} \citep{freeman2013new}, \emph{Gini}, $M_{20}$ \citep{lotz2004new}, \emph{C} and \emph{A} \citep{conselice2003relationship}. Each statistic (see the references for details) explains particular aspects of galaxy morphology.  In brief, the \emph{M, I, D} statistics capture galaxies with disturbed morphologies, \emph{Gini} and $M_{20}$ describe the variance of a galaxy's stellar light distribution, and  the \emph{C} and \emph{A} statistics measure the concentration of light and asymmetry of a galaxy, respectively. We restrict our study to relatively nearby galaxy observations that have a redshift (proxy for distance) estimate between $0.56 < z < 1.12$. The final data set consists of 2736 so-called $i$-band-selected galaxy observations. For each galaxy, we have seven morphological image statistics along with an estimate of star-formation rate (SFR).

Galaxy morphology is closely related to other physical properties such as star formation rate, mass and metallicity \citep[see, e.g.,][]{snyder2015galaxy}. The aim of this study is to demonstrate that our local two-sample framework can be valuable in detecting and quantifying dependencies between variables of moderate or high dimension without resorting to low-dimensional projections of summary statistics. In particular, we demonstrate that local two-sample tests can identify galaxies that lie in regions of the feature space where the estimated proportion of a particular defined class of objects (such as star-forming galaxies) differs significantly from the global proportion. Hence, we start by defining two galaxy classes based on the SFR: we say that a galaxy belongs to the high-SFR group if its SFR is higher than the upper 25\% quantile of the SFR distribution ($\log_{10}(\text{SFR})>1.201$), and that it belongs to the low-SFR group if its SFR is lower than the lower 25\% quantile of the SFR distribution ($\log_{10}(\text{SFR})<-0.915$). We further randomly divide the  data into a training set ($n=684$) and a test set ($n=684$). We use the training data to construct the local test statistic in (\ref{Eq: Test Statistic}), and we perform the local-two sample tests at the points in the test set (that is, these are the evaluation points in Algorithm~\ref{Alg: Local Two-Sample Testing via Permutations}).  Note that this particular application is especially challenging because the seven morphological statistics have very different properties, and some of the statistics ($M$ and $I$) are essentially of mixed discrete and continuous type with heavy outliers; hence, any metric-based estimator is bound to perform poorly even after normalizing the variables. Our regression test, however, can by-pass this problem by leveraging the random forest algorithm. Another advantage of using random forests is that the algorithm returns variable importance measures that can help us identify {\em which} morphology statistics are the most important in distinguishing the two populations (Figure~\ref{Fig: variable importance measure}).

\subsection{Analysis and Result}

\begin{figure}[t!]
	\begin{center}
		\includegraphics[width=7cm]{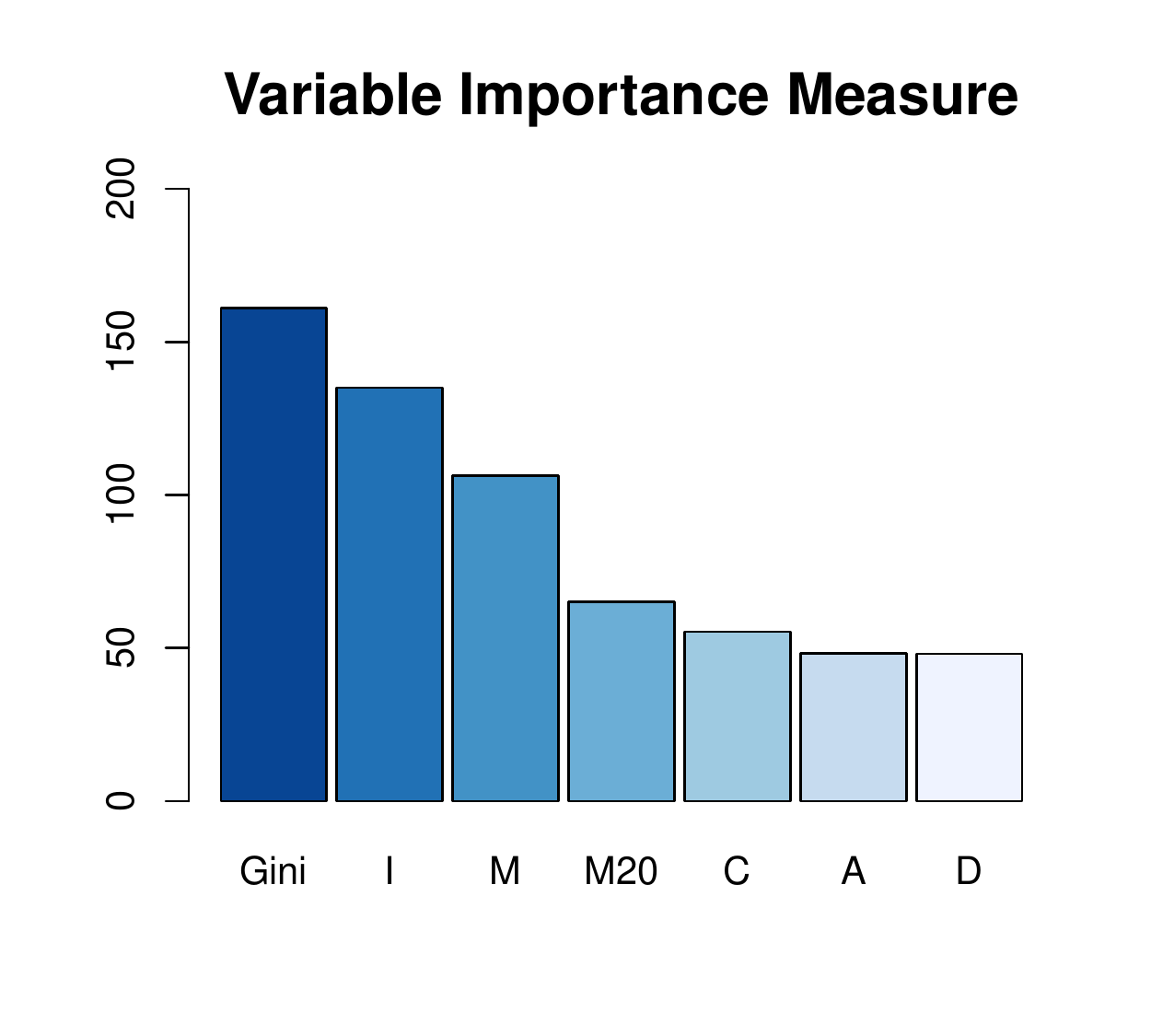}
		\caption{Variable importance measures from random forest regression, as measured by the Mean Decrease Gini (MDG) metric when splitting the data along the indicated variables. For the morphology-SFR study, the $Gini$ and $I$ morphology statistics are the two most important features in distinguishing between high-star-forming and the low-star-forming galaxy populations.} \label{Fig: variable importance measure}
	\end{center}
\end{figure}

According to our global two-sample test ($\widehat{\mT}_{RF}=.188$, $p<.001$), there is a significant difference between the low-SFR and the high-SFR populations in terms of galaxy morphology. We follow up on this result by implementing the local two-sample testing framework according to Algorithm \ref{Alg: Local Two-Sample Testing via Permutations} with FWER control at $\alpha =0.05$ by the Hochberg step up procedure. To visualize locally significant points from the local test, we use diffusion maps with local scaling \citep{zelnik2005self}. For more information on our particular application of diffusion maps, see Appendix~\ref{Sec: Diffusion Map}. The main result of the local significance test is displayed in Figure~\ref{Figure: Diffusion Map}. As we can see, the high-SFR and low-SFR dominated regions (that is, the regions where $f_{\text{LowSFR}}<f_{\text{HighSFR}}$ and $f_{\text{LowSFR}}>f_{\text{HighSFR}}$, respectively) are fairly well-separated in morphology space. Figure~\ref{Figure: Diffusion Map} also shows some examples of galaxy images at significant test points. By inspecting such images, we note that the ``red'' galaxies in the low-SFR dominated regions of the seven-dimensional space
tend to be more concentrated and less disturbed than their ``blue'' counterparts in the high-SFR dominated regions --- this result is consistent with previous astronomical studies about irregular galaxies displaying merger activities and high star-formation rates. Our test result is further supported by the variable importance measures in Figure~\ref{Fig: variable importance measure}: the two most important morphology statistics in distinguishing between high-SFR and low-SFR galaxies are the $Gini$ \citep{lotz2004new} and $I$ \citep{freeman2013new} morphology statistics. Indeed, by definition, the $Gini$ statistic describes the variance of a galaxy's stellar light distribution, and the $I$ statistic captures galaxies with disturbed morphologies.

\section{Conclusions} \label{Section: Conclusions} 
In this work, we presented a new framework for both global and local two-sample testing via regression. Depending on the chosen regression model, our framework can efficiently deal with different types of variables and different structures in the data; thereby, providing tests with competitive power against many practical alternatives. Compared to other recent approaches in the two-sample literature (such as classification tests), our framework has the key advantage of being able to detect locally significant regions in multivariate spaces. Throughout this work, we studied theoretical properties of the regression tests by building on existing regression results. We established a connection between the power of the global and local tests to the MISE and MSE of the corresponding regression estimators, and we demonstrated practical usefulness of our methods via simulations. 

By taking advantage of permutation tests under the global null hypothesis, the proposed local testing framework ensures that the type I error rate is less than or equal to the significance level. When the local null hypothesis $H_0(x): m(x) = \pi$ is of interest, on the other hand, there is no such guarantee. In this case, it would be necessary to use an asymptotic framework and investigate the limiting behavior of a local test statistic. This topic is reserved for future work. Another direction for future work is to study the optimality of global regression tests.  Contrary to the local regression test, a regression estimator with the optimal estimation error rate may not necessarily return minimax optimal global regression test. We hope that future studies will establish a lower bound and matching upper bound for the global regression test.

\addtocontents{toc}{\protect\setcounter{tocdepth}{1}}

\section*{Acknowledgements}

ABL would like to thank Rafael Izbicki and Larry Wasserman for discussions that lead to the two-sample testing work, and Peter Freeman and Jeffrey Newman for acting as IK's co-advisors for the data analysis project on which Section~\ref{Section: Application to Astronomy Data} is based. The authors also thank the editor and the reviewers for their constructive comments and suggestions. This work was partially supported by NSF DMS-1520786.

\bibliographystyle{apalike}
\bibliography{reference}


\appendix

\section{Proofs} \label{Section: Proofs}

\allowdisplaybreaks

%
%

\subsection{Proof of Theorem \ref{Theorem: Asymptotic null distribution of Fisher's LDA}} 
We start by simplifying $\widehat{m}_{\text{LDA}}(x)$ as
\begin{align*}
& \widehat{m}_{\text{LDA}}(X_i) \\[.5em]
=~ & \frac{ \pi_1 \exp \big\{ -\frac{1}{2}(X_i-\widehat{\mu}_1)^\top \mathcal{S}^{-1} (X_i- \widehat{\mu}_1) \big\} }{ \pi_1 \exp \big\{ -\frac{1}{2}(X_i- \widehat{\mu}_1)^\top \mathcal{S}^{-1} (X_i- \widehat{\mu}_1)  \big\} + \pi_0 \exp\big\{ -\frac{1}{2}(X_i- \widehat{\mu}_0)^\top \mathcal{S}^{-1} (X_i-\widehat{\mu}_0)  \big\}} \\[.5em]
=~ & \frac{ \pi_1 }{ \pi_1 + \pi_0 \exp \big\{ -\frac{1}{2}(X_i- \widehat{\mu}_0)^\top \mathcal{S}^{-1} (X_i- \widehat{\mu}_0) + \frac{1}{2}(X_i-\widehat{\mu}_1)^\top \mathcal{S}^{-1} (X_i- \widehat{\mu}_1) \big\}} \\[.5em]
=~ & \frac{ \pi_1}{ \pi_1 +  \pi_0\exp \Big\{ \left( X_i - ( \widehat{\mu}_0 + \widehat{\mu}_1) /2 \right)^\top \mathcal{S}^{-1} \left( \widehat{\mu}_0 - \widehat{\mu}_1 \right)   \Big\}},
\end{align*}
and write
\begin{align*}
W_i = \left( X_i - ( \widehat{\mu}_0 + \widehat{\mu}_1)/2 \right)^\top \mathcal{S}^{-1} \left( \widehat{\mu}_0 - \widehat{\mu}_1 \right).
\end{align*}
For some $a \in (0,1)$, Taylor expansion of $f(x) = a/\{ a+(1-a)e^{x}\}$ at $x=0$ provides
\begin{align*}
\big| \big\{ \widehat{m}_{\text{LDA}}(X_i) - \pi_1  \big\}^2 - \pi_0^2 \pi_1^2 W_i^2 \big|  \leq C |W_i|^3, 
\end{align*}
where $C$ is a universal constant. This implies that
\begin{align*}
\Bigg| \sum_{i=1}^n \Big\{ \widehat{m}_{\text{LDA}}(X_i) - \pi_1 \Big\}^2 -  \pi_0^2 \pi_1^2  \sum_{i=1}^n W_i^2    \Bigg| ~ \leq ~ C \sum_{i=1}^n |W_i|^3.
\end{align*}
Now based on $|x+y|^3 \leq 4 |x|^3 + 4 |y|^3$ and Cauchy-Schwarz inequality, it can be seen that 
\begin{align*}
\sum_{i=1}^n |W_i|^3  \leq 4n \big| ((\widehat{\mu}_0 + \widehat{\mu}_1)/2)^\top \mathcal{S}^{-1} (\widehat{\mu}_0 - \widehat{\mu}_1 )  \big|^3 + 4 \sum_{i=1}^n \big|X_i^\top \mathcal{S}^{-1} (\widehat{\mu}_0 - \widehat{\mu}_1) \big|^3 = o_P(1).
\end{align*}
As a result, $n\widehat{\mathcal{T}}_{\text{LDA}}$ can be approximated by 
\begin{align} 
n\widehat{\mathcal{T}}_{\text{LDA}}  =  \sum_{i=1}^n \Big\{ \widehat{m}_{\text{LDA}}(X_i) - \pi_1 \Big\}^2 = \pi_0^2 \pi_1^2  \sum_{i=1}^n W_i^2 + o_P(1). \label{TaylorRemainder}
\end{align}
Let us denote $\delta_n = \mathcal{S}^{-1} (\widehat{\mu}_0 - \widehat{\mu}_1)$ and $\Delta_n = (\widehat{\mu}_0 + \widehat{\mu}_1)/2$, and recall $\mathcal{S} = n^{-1} \sum_{i=1}^n (X_i - \widehat{\mu}) (X_i - \widehat{\mu})^\top$ where $\widehat{\mu} = n^{-1} \sum_{i=1}^n X_i$. Then we observe that
\begin{align*}
\frac{1}{n} \sum_{i=1}^n  W_i^2  & ~=~ \frac{1}{n} \sum_{i=1}^n \Big\{ \delta_n^\top X_i - \delta_n^\top \Delta_n  \Big\}^2 \\[.5em]
& ~=~ \delta_n^\top \Bigg\{  \frac{1}{n} \sum_{i=1}^n \left( X_i - \Delta_n \right) \left( X_i - \Delta_n \right)^\top  \Bigg\} \delta_n \\[.5em]
& ~=~ \delta_n^\top \mathcal{S} \delta_n + \delta_n^\top \left( \widehat{\mu} - \Delta_n \right)   \left( \widehat{\mu} - \Delta_n \right)^\top \delta_n  \\[.5em]
& ~ = ~ (\widehat{\mu}_0 - \widehat{\mu}_1)^\top \mathcal{S}^{-1}  (\widehat{\mu}_0 - \widehat{\mu}_1) + R_n,
\end{align*}
where $R_n= \delta_n^\top \left( \widehat{\mu} - \Delta_n \right)   \left( \widehat{\mu} - \Delta_n \right)^\top \delta_n$. Hence, we have
\begin{align*}
n\widehat{\mathcal{T}}_{\text{LDA}} & = n \pi_0^2 \pi_1^2 \Big\{ (\widehat{\mu}_0 - \widehat{\mu}_1)^\top \mathcal{S}^{-1} (\widehat{\mu}_0 - \widehat{\mu}_1) + R_n   \Big\}+ o_P(1).
\end{align*}
We also note that the residual term is negligible under the null, i.e. $n \pi_0^2 \pi_1^2 R_n = o_P(1)$, which results in
\begin{align*}
n \pi_0^{-1} \pi_1^{-1}\widehat{\mathcal{T}}_{\text{LDA}} & = \frac{n_0 n_1}{n_0 + n_1} (\widehat{\mu}_0 - \widehat{\mu}_1)^\top \mathcal{S}^{-1} (\widehat{\mu}_0 - \widehat{\mu}_1) + o_P(1)   \\[.5em] 
& = T_{\text{Hotelling}}^2 + o_P(1).
\end{align*}
The rest of the proof follows by the limiting property of Hotelling's $T^2$.

\subsection{Proof of Theorem~\ref{Theorem: Asymptotic distribution of LDA against local alternatives}}
\begin{proof}
	First note that the likelihood ratio for testing (\ref{Eq: localAlternative}) is given by
	\begin{align*}
	\mathcal{L}_{n} = \sum_{i=1}^{n_1} \log  \frac{f_{\mu_0 + h/\sqrt{n}}(X_{i,1}) }{f_{\mu_0}(X_{i,1})}.
	\end{align*}	
	Since $\{\mP_\mu, \mu \in \Omega\}$ is q.m.d.~at $\mu_0$, Theorem 12.2.3 of \cite{lehmann2006testing} under $n_1 /(n_0 + n_1) \rightarrow \pi_1$ yields that 
	\begin{align*}
	\mathcal{L}_{n} \convD N \left( -\frac{\pi_1}{2} \langle h, I(\mu_0)h \rangle , \ \pi_1 \langle  h, I(\mu_0)h \rangle \right),
	\end{align*}
	where $I(\mu)$ is the Fisher information matrix. This implies by Corollary 12.3.1 of \cite{lehmann2006testing} that the joint distribution of $X_{1,0}$ and $X_{1,1}$ under the null and the alternative are mutually contiguous. Since contiguity implies 
	\begin{align*}
	n{\pi_0^{-1} \pi_1^{-1}}\widehat{\mathcal{T}}_{\text{LDA}} & = \frac{n_0n_1}{n_0 + n_1} (\widehat{\mu}_0 - \widehat{\mu}_1)^\top \mathcal{S}^{-1} (\widehat{\mu}_0 - \widehat{\mu}_1) + o_P(1),
	\end{align*}
	under $H_{1,n}$, the result follows by the limiting distribution of Hotelling's $T^2$ statistic. 
\end{proof}

\subsection{Proof of Theorem~\ref{Theorem: General Results of Global Regression Tests}}
\begin{proof}
	The exact type I error control of the permutation test is well-known \citep[see e.g. Chapter 15 of][]{lehmann2006testing}. Strictly speaking, the considered test is not the usual permutation test since the only first half of labels are permuted to decide a critical value. However, it also controls the type I error under $H_0$ due to Theorem 15.2.1 of \cite{lehmann2006testing}. Indeed, this result holds regardless of i.i.d.~sampling or separate sampling. Hence we focus on the type II error control. 
	
	\vskip .8em 
	
	\noindent \textbf{$\bullet$ Type II error control (i.i.d.~sampling)}
	
	\vskip .5em
	
	\noindent We start with the case of i.i.d.~sampling. Based on the inequality $(x-y)^2 \leq 2 (x-z)^2 + 2 (z-y)^2$, we lower bound the test statistic as 
	\begin{align} \nonumber
	\widehat{\mathcal{T}}_{global}^\prime  & = \frac{1}{n} \sum_{i=n+1}^{2n} \left( \widehat{m}(X_i) - \widehat{\pi}_1 \right)^2 \\[.5em] \nonumber
	& \geq  \frac{1}{2n} \sum_{i=n+1}^{2n} \left( m(X_i) - \widehat{\pi}_1 \right)^2 - \frac{1}{n} \sum_{i=n+1}^{2n} \left( \widehat{m}(X_i) - m(X_i) \right)^2  \\[.5em] 
	& \geq \frac{1}{4n} \sum_{i=n+1}^{2n} \left( m(X_i) - \pi_1 \right)^2 - \frac{1}{2}(\pi_1 - \widehat{\pi}_1)^2 - \frac{1}{n} \sum_{i=n+1}^{2n} \left( \widehat{m}(X_i) - m(X_i) \right)^2. \label{Eq: Lower Bound of Statistic}
	\end{align}
	Define the events $\mathcal{A}_1, \mathcal{A}_2, \mathcal{A}_3, \mathcal{A}_4$ such that 
	\begin{align*}
	& \mathcal{A}_1 = \Big\{ (\pi_1 - \widehat{\pi}_1)^2 < C_2 \delta_n  \Big\}, \\[.5em]
	& \mathcal{A}_2 =  \Big\{  \frac{1}{n} \sum_{i=n+1}^{2n} \left( \widehat{m}(X_i) - m(X_i) \right)^2 < C_3\delta_n \Big\}, \\[.5em]
	&  \mathcal{A}_3 =  \Big\{  \Big|  \frac{1}{n} \sum_{i=n+1}^{2n} \left( m(X_i) - \pi_1 \right)^2 - \mE \left[ (m(X_i)- \pi_1)^2 \right] \Big| < \frac{1}{2} \mE \left[ (m(X)- \pi_1)^2 \right]  \Big\}, \\[.5em]
	& \mathcal{A}_4 =  \Big\{ t_\alpha < C_{0,\alpha}^\prime \delta_n \Big\}.
	\end{align*}
	Using Markov's inequality, we have 
	\begin{align*}
	& \mP\left( \mathcal{A}_1^c  \right) \leq \frac{\pi_1(1-\pi_1)}{C_2 n \delta_n}, \\[.5em]
	& \mP\left( \mathcal{A}_2^c  \right) \leq \frac{1}{C_3 \delta_n} \mE\left[ \int_S (\widehat{m}(x) -m(x) )^2 dP_X(x) \right] \leq \frac{C_0}{C_3},
	\end{align*}
	by the condition in (\ref{Eq: MSE}). For the third event, denote $\Delta_n =  \mE\left[ (m(X) - \pi_1 )^2 \right]$ and use Chebyshev's inequality to have
	\begin{align*}
	\mP \left( \mathcal{A}_3^c \right) & \leq \frac{4}{n\Delta_n^2}  \text{Var} \left[ (m(X) - \pi_1)^2 \right] \\[.5em]
	& \leq \frac{4}{n\Delta_n^2}  \mE \left[ (m(X) - \pi_1)^4 \right] \\[.5em]
	& \leq  \frac{4}{n\Delta_n^2}  \mE \left[ (m(X) - \pi_1)^2 \right]  \quad \text{since $|m(X) - \pi_1| \leq 1$ }\\[.5em]
	& \leq \frac{4}{C_1 n \delta_n},
	\end{align*}
	where the last inequality uses the assumption that $\Delta_n \geq C_1 \delta_n$. Furthermore, under the assumption on the permutation critical value, $\mP \left( \mathcal{A}_4^c \right) \leq \beta / 2$. Hence, we obtain
	\begin{align*}
	\mP \left( (\mathcal{A}_1 \cap \mathcal{A}_2 \cap \mathcal{A}_3 \cap \mathcal{A}_4)^c \right)  \leq \sum_{i=1}^4 \mP \left(\mathcal{A}_i^c \right) < \beta,
	\end{align*}
	by choosing sufficiently large $C_1,C_2,C_3 > 0$ with the assumption that $\delta_n \geq n^{-1}$. Using (\ref{Eq: Lower Bound of Statistic}), the type II error of the regression test is bounded by 
	\begin{align*}
	& \mP( \widehat{\mathcal{T}}_{global}^\prime  \leq  t_\alpha)  \\[.5em]
	\leq ~ & \mP \Bigg(  \frac{1}{4n} \sum_{i=n+1}^{2n} \left( m(X_i) - \pi_1 \right)^2 - \frac{1}{2}(\pi_1 - \widehat{\pi}_1)^2 - \frac{1}{n} \sum_{i=n+1}^{2n} \left( \widehat{m}(X_i) - m(X_i) \right)^2 \leq  t_\alpha \Bigg) \\[.5em]
	\leq ~ & \mP \Bigg( \Bigg\{ \frac{1}{4n} \sum_{i=n+1}^{2n} \left( m(X_i) - \pi_1 \right)^2 - \frac{1}{2}(\pi_1 - \widehat{\pi}_1)^2 - \frac{1}{n} \sum_{i=n+1}^{2n} \left( \widehat{m}(X_i) - m(X_i) \right)^2 \leq  t_\alpha \Bigg\}  \\[.5em] 
	& ~~~~~ \bigcap\Bigg\{ \bigcap_{j=1}^4 \mathcal{A}_j \Bigg\} \Bigg) + \mP \left( (\mathcal{A}_1 \cap \mathcal{A}_2 \cap \mathcal{A}_3 \cap \mathcal{A}_4)^c \right)  \\[.5em]
	\leq ~ & \mP \left( \Delta_n < C_4 \delta_n \right) + \beta,
	\end{align*}
	where $C_4$ can be chosen by $C_4 = 2C_{0,\alpha}^\prime + C_2 + 2C_3$. Now by choosing $C_1 > C_4$ for sufficiently large $n$, the type II error can be bounded by $\beta$. Hence the result follows.
	
	\vskip 1em 
	
	\noindent \textbf{$\bullet$ Type II error control (Separate Sampling)}
	
	\vskip .5em
	
	\noindent The proof for separate sampling is almost the same as before except few details. First, we do not need to define $\mathcal{A}_1$ since $\pi_1$ is known. In terms of $\mathcal{A}_2$, apply Markov's inequality to obtain 
	\begin{align*}
	\mP\left( \mathcal{A}_2^c  \right) & ~\leq~ \frac{1}{C_3 \delta_n} \Bigg\{ \frac{n_0}{n}\mE\left[ \int_S (\widehat{m}(x) -m(x) )^2 dP_0(x) \right] \\[.5em]
	&~~~~~~~~~~~~ + \frac{n_1}{n} \mE\left[ \int_S (\widehat{m}(x) -m(x) )^2 dP_1(x) \right] \Bigg\}  \\[.5em]
	& ~=~  \frac{C_0}{C_3}  \mE\left[ \int_S (\widehat{m}(x) -m(x) )^2 dP_X(x) \right] \leq \frac{C_0}{C_3},
	\end{align*} 
	where the last line uses the fact that $\frac{n_0}{n} P_0 + \frac{n_1}{n} P_1 = P_X$. Similarly, for the event $\mathcal{A}_3$, we have by Chebyshev's inequality that 
	\begin{align*}
	\mP \left( \mathcal{A}_3^c \right) & \leq \frac{4}{\Delta_n^2} \frac{1}{n^2}  \sum_{i=n+1}^{2n} \text{Var} \left[ (m(X_i) - \pi_1)^2 \right] \\[.5em]
	& \leq \frac{4}{\Delta_n^2} \frac{1}{n^2}  \sum_{i=n+1}^{2n} \mE \left[ (m(X_i) - \pi_1)^2 \right] =  \frac{4}{n\Delta_n^2}  \mE \left[ (m(X) - \pi_1)^2 \right]\\[.5em]
	& \leq \frac{4}{C_1 n \delta_n}.
	\end{align*}
	The rest follows exactly the same as before. Hence the proof is complete. 
\end{proof}

\subsection{Proof of Corollary~\ref{Corollary: Global Regression Test}}
\begin{proof}
	We prove the corollary by showing that the conditions in Theorem~\ref{Theorem: General Results of Global Regression Tests} are satisfied. In particular, it suffices to verify that for fixed $\alpha \in (0,1)$ and $\beta \in (0,1-\alpha)$, there exists a positive constant $C_{0,\alpha}^\prime$ such that $\sup_{f_0,f_1 \in \mathcal{M}}\mP_{f_0,f_1} (t_\alpha < C_{0,\alpha}^\prime \delta_n) \geq 1 - \beta/2$. Then the rest of the proof proceeds the same as before. 
	
	\vskip .8em
	
	\noindent \textbf{$\bullet$ i.i.d.~sampling} 
	
	\vskip .5em
	
	\noindent To start with the case of i.i.d.~sampling, let $\eta= (\eta_1,\ldots,\eta_{n})^\top$ be a permutation of $\{1,\ldots,{n}\}$. Now conditioned on the data $\mathcal{X}_{2n} = \{(X_1,Y_1),\ldots, (X_{2n},Y_{2n}) \}$, we denote the probability and expectation under permutations by $\mP_\eta[\cdot] = \mP_\eta[\cdot|\mathcal{X}_{2n}]$ and $\mE_\eta[\cdot] = \mE_\eta[\cdot|\mathcal{X}_{2n}]$ respectively. Then by Markov's inequality
	\begin{align*}
	\mP_\eta\left( \widehat{\mathcal{T}}_{global}^\prime  \geq t \right) & = \mP_\eta \left( \frac{1}{n} \sum_{i=n+1}^{2n} \left( \widehat{m}_\eta(X_i) - \widehat{\pi}_1 \right)^2 \geq t \right) \\[.5em]
	& \leq \frac{1}{t n } \sum_{i=n+1}^{2n} \mE_\eta \left[ (\widehat{m}_\eta(X_i) - \widehat{\pi}_1)^2 \right], 
	\end{align*}
	where $\widehat{m}_\eta(x) = \sum_{i=1}^{n} w_i(x) Y_{\eta_i}$. Since $\sum_{i=1}^n w_i(x) = 1$ for any $x \in S$, 
	\begin{align*}
	\mE_\eta \left[ \widehat{m}_\eta (x) \right] = \sum_{i=1}^n w_i(x) \mE_{\eta} [Y_{\eta_i}] = \sum_{i=1}^n w_i(x) \widehat{\pi}_1 = \widehat{\pi}_1.
	\end{align*}
	Further note that 
	\begin{align}  \label{Eq: Squared error under permutations}
	\mE_\eta \left[ (\widehat{m}_\eta(x) - \widehat{\pi}_1)^2 \right] & = \sum_{i_1=1}^{n} \sum_{i_2=1}^{n} w_{i_1}(x) w_{i_2}(x) \mE_{\eta} \left[ (Y_{\eta_{i_1}} - \widehat{\pi}_1 ) (Y_{\eta_{i_2}} - \widehat{\pi}_1 )\right] \\[.5em] \nonumber
	& \leq \sum_{i=1}^n w_i^2(x) \mE_{\eta} \left[ (Y_{\eta_i} - \widehat{\pi}_1)^2 \right] \\[.5em] \nonumber
	& = \widehat{\pi}_1 (1 - \widehat{\pi}_1 ) \sum_{i=1}^n w_i^2(x) \\[.5em] 
	& \leq \frac{1}{4} \sum_{i=1}^n w_i^2(x), \label{Eq: Upper bound of permutation moment}
	\end{align} 
	where the first inequality uses $ \mE_{\eta} \left[ (Y_{\eta_{i_1}} - \widehat{\pi}_1 ) (Y_{\eta_{i_2}} - \widehat{\pi}_1 )\right]  \leq 0$ when $i_1 \neq i_2$.

	Note that the permutation samples are not $i.i.d.$ and thus in order to use the condition in (\ref{Eq: MSE}) which holds for $i.i.d.$ samples, we will associate the upper bound in (\ref{Eq: Upper bound of permutation moment}) with $i.i.d.$ samples. To do so,	let $(Y_1^\ast,\ldots, Y_n^\ast)$ be $i.i.d.$ Bernoulli random variables with parameter $p = 1/2$ independent of $\{X_1,\ldots,X_{2n}\}$. Then 
	\begin{align*}
	& \mE_{Y^\ast} \left[ (\widehat{m}(x) - 1/2)^2 | X_1, \ldots, X_{2n} \right] \\[.5em]
	= ~ &   \mE_{Y^\ast} \Big[ \big(\sum_{i=1}^n w_i(x) Y_i^\ast - 1/2 \big)^2 \big| X_1, \ldots, X_{2n} \Big]  \\[.5em]
	= ~ &   \mE_{Y^\ast} \Big[ \big(\sum_{i=1}^n w_i(x) (Y_i^\ast - 1/2 ) \big)^2 \big| X_1, \ldots, X_{2n} \Big]    \\[.5em]
	= ~ & \sum_{i_1=1}^{n} \sum_{i_2=1}^{n} w_{i_1}(x) w_{i_2}(x) \mE_{Y^\ast} [ (Y_{i_1}^\ast - 1/2) (Y_{i_2}^\ast - 1/2) ] \\[.5em]
	= ~ & \frac{1}{4} \sum_{i=1}^n w_i^2(x). 
	\end{align*}
	Therefore, we obtain
	\begin{align*}
	\mE_\eta \left[ (\widehat{m}_\eta(x) - \widehat{\pi}_1)^2 \right] & \leq ~ \mE_{Y^\ast} \left[ (\widehat{m}(x) - 1/2)^2 | X_1,\ldots,X_{2n} \right]
	\end{align*}
	which in turn implies that
	\begin{align*}
	\mP_\eta \left( \widehat{\mathcal{T}}_{global}^\prime  \geq t \right)  \leq  \frac{1}{t n } \sum_{i=n+1}^{2n}\mE_{Y^\ast} \left[ (\widehat{m}(X_i) - 1/2 )^2 | X_1,\ldots,X_{2n} \right]. 
	\end{align*}
	So the critical value of the permutation distribution is bounded by 
	\begin{align} \label{Eq: Upper bound of C.V.}
	t_{\alpha}^\ast \leq  \frac{1}{ \alpha n} \sum_{i=n+1}^{2n}\mE_{Y^\ast} \left[ (\widehat{m}(X_i) - 1/2 )^2 | X_1,\ldots,X_{2n} \right].
	\end{align}
	Now choose $C_{0,\alpha}^\prime$ such that $2C_0/(\alpha\beta) \leq C_{0,\alpha}^\prime$. Then based on the assumption in (\ref{Eq: MSE}) and Markov's inequality
	\begin{align*}
	&\sup_{f_0,f_1 \in \mathcal{M}} \mP_{f_0,f_1} \left( t_{\alpha}^\ast \geq C_{0,\alpha}^\prime  \delta_n  \right) \\[.5em]
	\leq ~ &\sup_{f_0,f_1 \in \mathcal{M}} \mP_{f_0,f_1} \left(  \frac{1}{ \alpha n} \sum_{i=n+1}^{2n}\mE_{Y^\ast} \left[ (\widehat{m}(X_i) - 1/2 )^2 | X_1,\ldots,X_{2n} \right] \geq C_{0,\alpha}^\prime \delta_n \right) \\[.5em]
	\leq ~ & \frac{C_0}{C_{0,\alpha}^\prime \alpha} \leq  \beta/2.
	\end{align*}
	Hence the proof completes. 
	
	\vskip .8em
	
	\noindent \textbf{$\bullet$ Separate Sampling} 
	
	\vskip .5em
	
	\noindent Let $Y_1^{\ast\ast},\ldots,Y_n^{\ast\ast}$ be Bernoulli random variables with parameter $\widehat{\pi}_1$ such that $\sum_{i=1}^n Y_i^{\ast\ast} = n\widehat{\pi}_1$ and they are independent of $X_1,\ldots,X_{2n}$. In the case of separate sampling, the proof follows similarly by noting that the right-hand side of (\ref{Eq: Squared error under permutations}) is the same as 
	\begin{align*}
	& \sum_{i_1=1}^{n} \sum_{i_2=1}^{n} w_{i_1}(x) w_{i_2}(x) \mE_{\eta} \left[ (Y_{\eta_{i_1}} - \widehat{\pi}_1 ) (Y_{\eta_{i_2}} - \widehat{\pi}_1 )\right] \\[.5em]
	= ~ & \sum_{i_1=1}^{n} \sum_{i_2=1}^{n} w_{i_1}(x) w_{i_2}(x) \mE_{Y^{\ast\ast}} \left[ (Y_{i_1}^{\ast\ast} - \widehat{\pi}_1 ) (Y_{i_2}^{\ast\ast}- \widehat{\pi}_1 )\right] \\[.5em]
	= ~ & \mE_{Y^{\ast\ast}} [ (\widehat{m}(x)  - \widehat{\pi}_1)^2 |X_1,\ldots,X_n].
	\end{align*}
	Now by putting the above quantity into the right-hand side of (\ref{Eq: Upper bound of C.V.}) and following the same lines afterwards, we complete the proof. 
\end{proof}

\subsection{Proof of Theorem~\ref{Theorem: Local Regression Test}}
This result can be proved by following the same steps in the proof of Theorem~\ref{Theorem: General Results of Global Regression Tests}. 
In fact, it is simpler than the previous proof since it does not involve sample splitting to estimate the integration error; hence we omit the proof.

\subsection{Proof of Example~\ref{Example: kNN MSE}} 
\begin{proof}
	Let $\overline{m}_{kNN}(x) = \mE [ \widehat{m}_{kNN}(x) |X_1, \ldots, X_n ]$. Then we have the following decomposition.
	\begin{align*}
	\mE \left[ \left( \widehat{m}_{kNN}(x) - m(x) \right)^2 \right] = \underbrace{\mE \left[ \left( \widehat{m}_{kNN}(x) - \overline{m}_{kNN}(x) \right)^2 \right]}_{(I)} + \underbrace{\mE \left[ \left( \overline{m}_{kNN}(x) - m(x) \right)^2 \right]}_{(II)}.
	\end{align*}
	For a fixed $x$, Proposition 8.1 of \cite{biau2015lectures} shows that conditioned on $\{X_1,\ldots,X_n\}$, 
	\begin{align*}
	(X_{1,n}(x),Y_{1,n}(x)), \ldots, ((X_{n,n}(x),Y_{n,n}(x)))
	\end{align*}
	are independent. Using this independence property,  
	\begin{align*}
	(I)= ~   \mE \left[  \left( \frac{1}{k_n} \sum_{i=1}^{k_n} (Y_{i,n}(x) - m(X_{i,n}(x) )) \right)^2  \right]  \leq ~  \frac{1}{4k_n}.
	\end{align*}
	Next for $(II)$, 
	\begin{align*}
	(II) = ~ & \mE \left[  \left( \frac{1}{k_n} \sum_{i=1}^{k_n} (m(X_{i,n}(x) )  - m(x) )\right)^2  \right]  \\[.5em] 
	\leq ~ &  \mE \left[  \left( \frac{1}{k_n} \sum_{i=1}^{k_n} \big| m(X_{i,n}(x) )  - m(x) \big| \right)^2  \right]  \\[.5em] 
	\leq ~ &  \mE \left[  \left( \frac{L}{k_n} \sum_{i=1}^{k_n} || X_{i,n}(x)  - x ||_2 \right)^2  \right]
	\end{align*}
	where the last inequality uses the Lipschitz condition. Note that for fixed $\epsilon >0$
	\begin{equation}
	\begin{aligned} \label{Eq: kNN measure1}
	\mP \left( || X_{1,n}(x) - x||_2 > \epsilon \right) & = \left( 1 - \mP(X \in B_{x,\epsilon}) \right)^n \\[.5em]
	& \leq  (1 - \tau_x \epsilon^{D})^n \leq ~ e^{-\tau_x n \epsilon^D}
	\end{aligned} 
	\end{equation}
	by the assumption that $\mP(X \in B_{x,\epsilon})  > \tau_x \epsilon^{D}$. Hence,
	\begin{align} \nonumber
	\mE \left[ || X_{1,n}(x) - x||^2 \right] =  &  \int_0^\infty \mP\left( || X_{1,n}(x) - x||_2 > \sqrt{\epsilon}  \right) d\epsilon \\[.5em] \nonumber 
	\leq  & \int_0^\infty  e^{-\tau_x n \epsilon^{D/2}} d\epsilon \\[.5em]  \label{Eq: Upper bound of the first nearest neighbor}
	=  & \frac{2\Gamma(2/D)}{D \tau_x^{2/D}} n^{-2/D}.
	\end{align}
	Similarly to the proof of Theorem 6.2 of \cite{gyorfi2002distribution}, divide the data into $k_n + 1$ parts where the first $k_n$ parts have size $\lfloor n/k_n \rfloor$ and denote the first nearest neighbor of $x$ from the $j$th partition by $\widetilde{X}_j^{x}$. This implies that
	\begin{align*}
	\sum_{i=1}^{k_n} || X_{i,n}(x)  - x ||_2 \leq \sum_{i=1}^{k_n} || \widetilde{X}_{i}^{x} - x ||_2 
	\end{align*} 
	and by Jensen's inequality,
	\begin{align*}
	(II) & \leq ~ \mE \left[  \left( \frac{L}{k_n} \sum_{i=1}^{k_n} || \widetilde{X}_{i}^{x} - x ||_2  \right)^2  \right] \\[.5em]
	& \leq ~  \frac{L^2}{k_n} \sum_{i=1}^{k_n} \mE \left[ || \widetilde{X}_{i}^{x} - x ||_2^2 \right] \\[.5em]
	& \leq ~ L^2 \frac{2\Gamma(2/D)}{D \tau_x^{2/D}} \left(\frac{k_n}{n}\right)^{2/D}
	\end{align*}
	by (\ref{Eq: Upper bound of the first nearest neighbor}). Combining the results, we have
	\begin{align*}
	\mE \left[ \left( \widehat{m}_{kNN}(x) - m(x) \right)^2 \right]  & = ~ (I) + (II) \\[.5em]
	& \leq ~ \frac{1}{4k_n} +  L^2 \frac{2\Gamma(2/D)}{D \tau_x^{2/D}} \left(\frac{k_n}{n}\right)^{2/D}.
	\end{align*}
	This completes the proof.
\end{proof}

\subsection{Proof of Example~\ref{Example: kernel MSE}}
\begin{proof}
	Following the proof of Example~\ref{Example: kNN MSE}, let 
	\begin{align*}
	\overline{m}_{ker}(x) = \mE \left[ \widehat{m}_{ker}(x) | X_1,\ldots,X_n \right]
	\end{align*}
	and thus
	\begin{align*}
	\mE \left[ \left( \widehat{m}_{ker}(x) - m(x) \right)^2 \right] = \underbrace{\mE \left[ \left( \widehat{m}_{ker}(x) - \overline{m}_{ker}(x) \right)^2 \right]}_{(I)} + \underbrace{\mE \left[ \left( \overline{m}_{ker}(x) - m(x) \right)^2 \right]}_{(II)}.
	\end{align*}
	Define an event 
	\begin{align*}
	\mathcal{A}_n = \Bigg\{ \sum_{i=1}^n K \left( \frac{x- X_i}{h_n} \right) \geq \lambda \Bigg\}.
	\end{align*}
	Then 
	\begin{align*}
	(I) & =~ \underbrace{\mE \left[ \left( \widehat{m}_{ker}(x) - \overline{m}_{ker}(x) \right)^2 I(\mathcal{A}_n) \right]}_{(I_1)} + \underbrace{\mE \left[ \left( \widehat{m}_{ker}(x) - \overline{m}_{ker}(x) \right)^2 I(\mathcal{A}_n^c) \right]}_{(I_2)}.
	\end{align*}
	For $(I_1)$, we have 
	\begin{align*}
	& \mE \left[  \left( \widehat{m}_{ker}(x) - \overline{m}_{ker}(x) \right)^2 I(\mathcal{A}_n)  | X_1,\ldots,X_n \right] \\[.5em]
	= ~ &\frac{\sum_{i=1}^n \text{Var}(Y_i|X_i)K \left( \frac{x-X_i}{h_n} \right) }{ \left( \sum_{i=1}^n K\left(\frac{x-X_i}{h_n}\right)  \right)^2 } I(\mathcal{A}_n)  \\[.5em]
	\leq ~ & \frac{1}{4\sum_{i=1}^n K\left(\frac{x-X_i}{h_n}\right)} I\left( \sum_{i=1}^n K\left(\frac{x-X_i}{h_n}\right) \geq \lambda \right) \\[.5em]
	\leq ~ & \frac{1+ \lambda^{-1}}{4 +4\sum_{i=1}^n K\left(\frac{x-X_i}{h_n}\right)} \\[.5em]
	\leq ~ & \frac{1+ \lambda^{-1}}{4 + 4\lambda \sum_{i=1}^n I\left( ||x - X_i||_2 \leq r h_n \right)} \\[.5em]
	\leq ~ & \frac{1 + \lambda}{4\lambda^2} \frac{1}{1 + \sum_{i=1}^n I\left( ||x - X_i||_2 \leq r h_n \right)}.
	\end{align*}
	By Lemma 4.1 of \cite{gyorfi2002distribution},
	\begin{align*}
	\mE\left[ \frac{1}{1 + B}\right] \leq \frac{1}{(n+1)p} \leq \frac{1}{np},
	\end{align*}
	where $B \sim \text{Binominal}(n,p)$. Using this result,
	\begin{align*}
	(I_1)  & \leq ~ \frac{1+\lambda}{4\lambda^2} \frac{1}{n \mP\left( X \in B_{x,rh_n} \right)} \\[.5em]
	& \leq ~ \left( \frac{1+\lambda}{4\lambda^2 \tau_x r^d} \right)  \frac{1}{n h_n^d}.
	\end{align*}
	
	\vskip .8em 
	
	For $(I_2)$, note that $\left( \widehat{m}_{ker}(x) - \overline{m}_{ker}(x) \right)^2 \leq 1$ and thus
	\begin{align*}
	(I_2)  & \leq ~ \mP \left( \sum_{i=1}^n K \left( \frac{x - X_i}{h_n} \right) < \lambda \right) \\[.5em]
	& \leq ~ \mP \left( \sum_{i=1}^n I \left( ||x - X_i||_2 \leq r h_n \right) = 0 \right)
	\end{align*}
	where the second inequality is because if there exists $X_i$ such that $||x - X_i||_2 \leq rh_n$, then we have $\sum_{i=1}^n K \left( \frac{x - X_i}{h_n} \right) \geq \lambda$ by the assumption on the kernel. In addition, 
	\begin{equation}
	\begin{aligned} \label{Eq: kernel measure1}
	\mP \left( \sum_{i=1}^n I \left( ||x - X_i||_2 \leq r h_n \right) = 0 \right) = ~ &  \left( 1 - \mP\left( X \in B_{x,rh_n} \right) \right)^n \\[.5em]
	\overset{(i)}{\leq} ~ & e^{- n \tau_x r^D h_n^D } \overset{(ii)}{\leq} ~ \left( \frac{e^{-1}}{\tau_x r^D} \right) \frac{1}{n h_n^D},
	\end{aligned}
	\end{equation}
	where $(i)$ uses $1 + x \leq e^x$ with the assumption $\mP\left( X \in B_{x,\epsilon} \right) \geq \tau_x \epsilon^D$ and $(ii)$ uses $\sup_z ze^{-z} \leq e^{-1}$. As a result,
	\begin{align*}
	(I) = (I_1) + (I_2) \leq \left( \frac{1+\lambda}{4\lambda^2 \tau_x r^D} +  \frac{e^{-1}}{\tau_x r^D}  \right)  \frac{1}{n h_n^D}.
	\end{align*}
	
	\vskip .8em
	
	For $(II)$, we use Jensen's inequality and the Lipschitz condition to have
	\begin{align*}
	& \left( \overline{m}_{ker}(x) - m_{ker}(x) \right)^2  \\
	= ~ &  \left(  \frac{\sum_{i=1}^n \left( m(X_i) - m(x) \right) K \left( \frac{x - X_i}{h_n}\right) }{\sum_{i=1}^n K \left( \frac{x - X_i}{h_n}\right) } \right)^2 I \left( \sum_{i=1}^n K \left( \frac{x - X_i}{h_n}\right)  > 0 \right) \\[.5em]
	& + m_{ker}(x)^2  I \left( \sum_{i=1}^n K \left( \frac{x - X_i}{h_n}\right)  = 0 \right)  \\[.5em]
	\leq ~ & \frac{\sum_{i=1}^{n} L^2 ||X_i - x||_2^2 K\left( \frac{x - X_i}{h_n} \right) }{\sum_{i=1}^n K \left( \frac{x- X_i}{h_n}\right)} I \left( \sum_{i=1}^n K \left( \frac{x - X_i}{h_n}\right)  > 0 \right) \\[.5em]
	& + I \left( \sum_{i=1}^n K \left( \frac{x - X_i}{h_n}\right)  = 0 \right).
	\end{align*}
	Since $K(x) \leq I(x \in B_{0,R})$, we observe that 
	\begin{align*}
	||X_i - x||_2^2 K \left( \frac{x- X_i}{h_n}\right) \leq R^2 h_n^2  K \left( \frac{x- X_i}{h_n}\right).
	\end{align*}
	Consequently,
	\begin{align*}
	\left( \overline{m}_{ker}(x) - m_{ker}(x) \right)^2 \leq ~ & L^2 R^2 h_n^2 + I \left( \sum_{i=1}^n K \left( \frac{x - X_i}{h_n}\right)  = 0 \right) \\[.5em]
	\leq ~ & L^2 R^2 h_n^2 + I \left( \sum_{i=1}^n I \left( ||x - X_i ||_2 \leq r h_n \right) = 0 \right),
	\end{align*}
	where the second inequality is by the assumption $ \lambda I(x \in B_{0,r}) \leq  K(x)$. By taking the expectation, 
	\begin{align} \label{Eq: kernel measure2}
	(II) \leq ~ & L^2 R^2 h_n^2 +  \left( 1 - \mP \left( X \in B_{x, r h_n} \right)  \right)^n \\[.5em]  \nonumber 
	\leq ~ & L^2 R^2 h_n^2 + \left( 1 - \tau_x r^D h_n^D \right)^n \\[.5em] \nonumber
	\leq ~ & L^2 R^2 h_n^2 + \left( \frac{e^{-1}}{\tau_x r^D} \right) \frac{1}{n h_n^D}. \nonumber
	\end{align}
	Therefore, we conclude that
	\begin{align*}
	\mE \left[ \left( \widehat{m}_{ker}(x) - m(x) \right)^2 \right] = ~ & (I) + (II) \\[.5em]
	\leq ~ & \left( \frac{1+\lambda}{4\lambda^2 \tau_x r^D} +  \frac{2 e^{-1}}{\tau_x r^D}  \right)\frac{1}{n h_n^D}  +    L^2 R^2 h_n^2,
	\end{align*}
	which completes the proof.
\end{proof}

\subsection{Proof of Theorem~\ref{Theorem: Lower Bound}}
\begin{proof}
	Suppose $X$ has the uniform distribution over $[0,B]^D$ and $B>0$. In addition, assume that for $0 < \epsilon < 1/2$, the regression function is given by
	\begin{equation} \label{Eq: m(x) setting}
	\begin{aligned}
	m(x) & = \epsilon \prod_{i=1}^D \left( 1 - \frac{x_i}{B\epsilon} \right)I\left(0 \leq x_i \leq B\epsilon \right) \\[.5em]
	& + \epsilon \prod_{i=1}^D \left( \frac{B(1-\epsilon)-x_i}{B\epsilon} \right) I \{B(1-\epsilon) \leq x_i \leq B \} + \frac{1}{2}
	\end{aligned}
	\end{equation}
	for $x = (x_1,\ldots,x_D) \in [0,B]^D$ and $m(x)=0$ otherwise. Therefore, we have $\pi_1 = \pi_0 = 1/2$. Now for any $x, z \in [0,B]^D$, the telescoping argument gives
	\begin{align*}
	& |m(x_1,\ldots,x_D) - m(z_1,\ldots,z_D)| \\[.5em]
	\leq ~ & |m(x_1,x_2,\ldots,x_D) - m(z_1,x_2,\ldots,x_D) | \\[.5em]
	+~ & \sum_{i=1}^{D-2} |m(z_1,\ldots, z_i, x_{i+1},\ldots ,x_D) - m(z_1,\ldots, z_i, z_{i+1}, x_{i+2}, \ldots ,x_D)|  \\[.5em]
	+~ & |m(z_1,z_2,\ldots,z_{D-1},x_D) - m(z_1,z_2,\ldots,z_D) |.
	\end{align*}
	For the first term,
	\begin{align*}
	& |m(x_1,x_2,\ldots,x_D) - m(z_1,x_2,\ldots,x_D) | \\[.5em]
	\leq ~ &  \epsilon \Bigg| \left(1 - \frac{x_1}{B\epsilon}\right) I\left(0 \leq x_1 \leq B\epsilon \right) - \left(1 - \frac{z_1}{B \epsilon}\right) I\left(0 \leq z_1 \leq B \epsilon \right)  \Bigg| \\[.5em]
	& \times \prod_{i=2}^D \Bigg| \left( 1 - \frac{x_i}{B \epsilon} \right)I\left(0 \leq x_i \leq B \epsilon \right) \Bigg| \\[.5em]
	+& \epsilon \Bigg|  \left( \frac{B(1-\epsilon)-x_1}{B\epsilon} \right) I\Big\{ B(1-\epsilon) \leq x_1 \leq B  \Big\} \\[.5em]
	&~~~~~~~~~~~~~~~~~~~~~~ - \left( \frac{B(1-\epsilon)-z_1}{B\epsilon} \right) I\Big\{ B(1-\epsilon) \leq z_1 \leq B  \Big\}  \Bigg|   \\[.5em] 
	& \times \prod_{i=2}^D \Bigg| \left( \frac{B(1-\epsilon)-x_i}{B\epsilon} \right) I\Big\{ B(1-\epsilon) \leq x_i \leq B  \Big\}   \Bigg| \\[.5em]
	\leq ~ &  \epsilon \Bigg| \left(1 - \frac{x_1}{B\epsilon}\right) I\left(0 \leq x_1 \leq B\epsilon \right) - \left(1 - \frac{z_1}{B \epsilon}\right) I\left(0 \leq z_1 \leq B \epsilon \right)  \Bigg| \\[.5em]
	+&  \epsilon \Bigg|  \left( \frac{B(1-\epsilon)-x_1}{B\epsilon} \right) I\Big\{ B(1-\epsilon) \leq x_1 \leq B  \Big\} \\[.5em]
	& ~~~~~~~~~~~~~~~~~~~~~~- \left( \frac{B(1-\epsilon)-z_1}{B\epsilon} \right) I\Big\{ B(1-\epsilon) \leq z_1 \leq B  \Big\}  \Bigg| \\[.5em]
	\leq ~ & \frac{2}{B} |x_1 - z_1| \leq \frac{2}{B} ||x - z||_2. 
	\end{align*}
	Applying the same logic to the other terms, we see that 
	\begin{align*}
	|m(x) - m(z)| \leq \frac{2D}{B} ||x-z||_2.
	\end{align*}
	By choosing $B = 2D/L$, the regression function $m(x)$ becomes $L$-Lipschitz with
	\begin{align*}
	\delta_{n,x} = |m(x) - \pi_1|^2 = \epsilon^2 \quad \text{at $x = (0,\ldots,0)$}.
	\end{align*}
	
	Next, we lower bound the testing error. Denote the product and joint measure of $(X,Y)$ described above by $P_0$ and $P_1$ respectively. Using the standard approach to lower bound the testing error \citep[e.g.][]{baraud2002non}, we obtain that for any $\alpha$ level test functions $\phi : \{ (X_1,Y_1), \ldots, (X_n,Y_n) \} \mapsto \{0,1\}$,
	\begin{align*}
	\inf_{\phi \in \Phi_{n,\alpha}} \sup_{f_0,f_1 \in \mathcal{M}_{Lip}(\delta_{n,x})} \mP_{f_0,f_1}(\phi = 0) \geq ~ 1 - \alpha - \text{TV}(P_0^n, P_1^n)
	\end{align*}
	where TV denotes total variation distance. Based on Pinsker's inequality, we get
	\begin{align*}
	\text{TV}(P_0^n, P_1^n) \leq  ~ \sqrt{\frac{n}{2} D_{KL} (P_1 || P_0)}
	\end{align*}
	where $D_{KL}$ is the Kullback-Leibler divergence and by the Jensen's inequality
	\begin{align*}
	& D_{KL} (P_1 || P_0) \\[.5em]
	= ~ &   \int \pi_1 f(x) \log \frac{f(x, Y=1)}{\pi_1 f(x)} dx +  \int (1-\pi_1) f(x) \log \frac{f(x, Y=0)}{(1- \pi_1) f(x)} dx \\[.5em]
	= ~ & \frac{1}{2} \int f(x) \log \frac{f(x|Y=1)}{f(x)} dx +  \frac{1}{2} \int f(x) \log \frac{f(x|Y=0)}{f(x)} dx \\[.5em]
	\leq ~ &  \frac{1}{2} \int \frac{(f(x|Y=1)-f(x) )^2}{f(x)}dx + \frac{1}{2} \int \frac{(f(x|Y=0)-f(x) )^2}{f(x)}dx.
	\end{align*}
	By the assumption on $(X,Y)$, $X$ has the marginal density $f(x) = B^{-D}$ and the conditional densities $f(x|Y=1) = 2B^{-D}m(x)$ and $f(x|Y=0) = 2B^{-D} - f(x|Y=1)$ for $x \in [0,B]^D$. Therefore,
	\begin{align*}
	& \frac{1}{2} \int \frac{(f(x|Y=1)-f(x) )^2}{f(x)}dx + \frac{1}{2} \int \frac{(f(x|Y=0)-f(x) )^2}{f(x)}dx \\[.5em]
	= ~ & \int \frac{(f(x|Y=1) - f(x))^2}{f(x)}dx \\[.5em]
	= ~ & 4 B^{-D} \int \left( m(x) - 1/2 \right)^2 dx.
	\end{align*}
	Using the definition of $m(x)$ in (\ref{Eq: m(x) setting}), the above integration is calculated by 
	\begin{align*}
	4 B^{-D} \int \left( m(x) - 1/2 \right)^2 dx = \frac{8}{3^D} \epsilon^{2+D}.
	\end{align*}
	Now by choosing $\epsilon = \beta^{2/(2+D)} 3^{D/(2+D)} 2^{-2/(2+D)}n^{-1/(2+D)}$, we have
	\begin{align*}
	\inf_{\phi \in \Phi_{n,\alpha}}  \sup_{f_0,f_1 \in \mathcal{M}_{Lip}(C_{1,x}n^{-2/(2+D)})} \mP_{f_0,f_1}(\phi = 0) \geq ~ 1 - \alpha - \beta.
	\end{align*}
	This completes the proof.
\end{proof}

\subsection{Proof of Proposition~\ref{Proposition: Local Testing Error of kNN and kernel for manifold data}}
It is enough to show that there exist universal constants $C_0,C_{0,\alpha}^\prime$ such that
\begin{align*}
\sup_{f_0,f_1 \in \mathcal{M}_{Lip}} \mE \left[ \left( \widehat{m}_{kNN}(x) - m(x) \right)^2 \right] \leq C_0n^{-\frac{2}{2+d}}, \\[.5em]
\sup_{f_0,f_1 \in \mathcal{M}_{Lip}} \mE \left[ \left( \widehat{m}_{ker}(x) - m(x) \right)^2 \right] \leq C_{0,\alpha}^\prime n^{-\frac{2}{2+d}}.
\end{align*}
Then we can apply Theorem~\ref{Theorem: Local Regression Test} to complete the proof. To start with kNN regression, we only need to modify (\ref{Eq: kNN measure1}) and follow the same steps in the proof of Example~\ref{Example: kNN MSE}. From the definition of the $(C,d)$-homogeneous measure, we see that
\begin{align*}
\mP \left( X \in B_{x,\epsilon} \right) \geq ~ \frac{\epsilon^{d}}{C} \mP\left( X \in B_{x,1} \right) = C^\prime \epsilon^{d}.
\end{align*}
As a result, (\ref{Eq: kNN measure1}) becomes
\begin{align*}
\mP \left( || X_{1,n}(x) - x||_2 > \epsilon \right) & = ~ \left( 1 - \mP(X \in B_{x,\epsilon}) \right)^n \\[.5em]
& \leq ~ \left( 1 -  C^\prime \epsilon^{d} \right)^n \leq e^{-C^\prime n\epsilon^{d}}.
\end{align*}
Then we end up having
\begin{align*}
\mE \left[ \left( \widehat{m}_{kNN}(x) - m(x) \right)^2 \right]   \leq ~ \frac{1}{4k_n} +  L^2 \frac{2\Gamma(2/{d})}{{d} C^{\prime 2/{d}}} \left(\frac{k_n}{n}\right)^{2/d}
\end{align*}
and the result follows by setting $k_n = n^{\frac{2}{2+d}}$. Similarly, we only need to modify (\ref{Eq: kernel measure1}) and (\ref{Eq: kernel measure2}) in the proof of Example~\ref{Example: kernel MSE}. By using the $(C,d)$-homogeneous measure, 
\begin{align*}
\left( 1 - \mP \left( X \in B_{x,rh_n} \right) \right)^n & \leq ~ \left( 1 - \frac{h_n^{d}}{C}\mP \left(X \in B_{x,r} \right) \right)^n \\[.5em]
& = ~\left( 1 - C^\prime h_n^{d} \right)^n \\[.5em]
& \leq ~ e^{-C^\prime n h_n^{d}}
\end{align*}
and apply this result to (\ref{Eq: kernel measure1}) and (\ref{Eq: kernel measure2}). We complete the proof by following the same steps in the proof of Example~\ref{Example: kernel MSE}.

\subsection{Proof of Theorem~\ref{Theorem: Limiting distribution of Local Regression Test}}
\begin{proof}
	We use a combinatorial central limit theorem in \cite{bolthausen1984estimate} to prove the result. First denote $a_{ij} = w_i(x) Y_j$ for $1\leq i,j \leq n$ and
	\begin{align*}
	\mu = n a_{\boldsymbol{\cdot} \boldsymbol{\cdot}}, \quad \sigma_n^2 = \sum_{1 \leq i,j \leq n}^n (a_{ij} - a_{i\boldsymbol{\cdot}}  - a_{\boldsymbol{\cdot} j} + a_{\boldsymbol{\cdot} \boldsymbol{\cdot}})^2 / (n-1),
	\end{align*}
	where
	\begin{align*}
	a_{i\boldsymbol{\cdot}} = \sum_{j=1}^n a_{ij} / n,  \quad a_{\boldsymbol{\cdot} j} = \sum_{i=1}^n a_{ij} / n, \quad a_{\boldsymbol{\cdot} \boldsymbol{\cdot}} = \sum_{1 \leq i,j \leq n}^n a_{ij} / n^2.
	\end{align*}
	In our case, $\mu = \widehat{\pi}_1$ and $\sigma_n^2$ is given in (\ref{Eq: Sigma Expression}). Let $d_{ij}= a_{ij} - a_{i\boldsymbol{\cdot}}  - a_{\boldsymbol{\cdot} j} + a_{\boldsymbol{\cdot} \boldsymbol{\cdot}} = (w_i(x) - 1/n) (Y_j - \widehat{\pi}_1)$. Then using the theorem in \cite{bolthausen1984estimate}, we obtain
	\begin{align*}
	\sup_{t \in \mathbb{R}} \bigg| \mP \left( \frac{\widehat{m}(x) - \widehat{\pi}_1}{\sigma_n} \leq t \Big| X_1,\ldots,X_n \right) - \Phi\left( t \right) \bigg|  \leq K \frac{1}{\sqrt{n}}  \frac{\frac{1}{n^2} \sum_{i,j} |d_{i,j}|^3 }{\left( \frac{1}{n^2} \sum_{i,j} d_{i,j}^2 \right)^{3/2} },
	\end{align*}
	where $K$ is a universal constant. Note that
	\begin{align*}
	\frac{1}{n^2} \sum_{i,j} |d_{i,j}|^3 & = ~ \frac{1}{n}\sum_{i=1}^n \Big| w_i(x) - \frac{1}{n} \Big|^3 \cdot \frac{1}{n} \sum_{j=1}^n \big| Y_j -\widehat{\pi}_1  \big|^3
	\end{align*}
	and 
	\begin{align*}
	\frac{1}{n^2} \sum_{i,j} d_{i,j}^2  = ~ \frac{1}{n}\sum_{i=1}^n \left( w_i(x) - \frac{1}{n} \right)^2 \cdot \frac{1}{n} \sum_{j=1}^n \left( Y_j -\widehat{\pi}_1  \right)^2.
	\end{align*}
	As a result,
	\begin{align*}
	\frac{\frac{1}{n^2} \sum_{i,j} |d_{i,j}|^3 }{\left( \frac{1}{n^2} \sum_{i,j} d_{i,j}^2 \right)^{3/2} }  & ~ = ~  \frac{1}{\sqrt{n}} \frac{\frac{1}{n} \sum_{i=1}^n \Big| w_i(x) - \frac{1}{n} \Big|^3}{ \Big\{ \frac{1}{n}\sum_{i=1}^n \Big( w_i(x) - \frac{1}{n} \Big)^{2} \Big\}^{3/2} }  \cdot \underbrace{\frac{\frac{1}{n} \sum_{j=1}^n \big| Y_j -\widehat{\pi}_1  \big|^3}{\left(\frac{1}{n} \sum_{j=1}^n \left( Y_j -\widehat{\pi}_1  \right)^2\right)^{3/2}}}_{(II)} \\[.5em]
	& ~ \leq ~ \underbrace{\frac{\max_{1 \leq i \leq n} (w_i(x) - 1/n)^2 }{\sum_{i=1}^n (w_i(x) - 1/n)^2}}_{(I)} \cdot (II)
	\end{align*}
	Note that $(I) = o_P(1)$ under the given assumption and $(II)$ is stochastically bounded by the law of large number. Thus we conclude that
	\begin{align*}
	\sup_{t \in \mathbb{R}} \bigg| \mP \left( \frac{\widehat{m}(x) - \widehat{\pi}_1}{\sigma_n} \leq t \Big| X_1,\ldots,X_n \right) - \Phi\left( t \right) \bigg| = o_P(1),
	\end{align*}
	which implies the desired result.
\end{proof}

\subsection{Proof of Corollary~\ref{Corollary: kNN example}}
\begin{proof}
	For kNN regression, there are $k$ and $(n-k)$ number of $k^{-1}$ and zero in $\{w_1(x),\ldots,w_n(x)\}$ respectively. Hence,
	\begin{align*}
	\sum_{i=1}^n \left( w_i(x) - \frac{1}{n} \right)^2 = k \left( \frac{1}{k} - \frac{1}{n} \right)^2 + \frac{n-k}{n^2}.
	\end{align*}
	Furthermore, under the assumption that $2k < n$, we have 
	\begin{align*}
	\max_{1 \leq i \leq n}  \bigg| w_i(x) -\frac{1}{n}  \bigg|  = \frac{1}{k} -\frac{1}{n}. 
	\end{align*}
	After direct calculations, one can show that
	\begin{align*}
	\frac{\max_{1 \leq i \leq n} |w_i(x) - 1/n|}{ \{ \sum_{i=1}^n (w_i(x) - 1/n)^2  \}^{1/2}} \rightarrow 0, 
	\end{align*}
	and thus the result follows.
\end{proof}	

\subsection{Proof of Corollary~\ref{Corollary: kernel example}}
\begin{proof}
	Note that
	\begin{align*}
	\widehat{m}_{ker}(x) = \sum_{i=1}^n w_i(x) Y_i = \frac{\sum_{i=1}^n Y_i K \left( \frac{x- X_i}{h_n} \right)}{\sum_{i=1}^n K \left( \frac{x- X_i}{h_n} \right)} = \frac{\sum_{i=1}^n Y_i K_{h_n} \left( x - X_i \right)}{\sum_{i=1}^n K_{h_n} \left( x - X_i \right)}.
	\end{align*}
	Hence it suffices to show that
	\begin{align*}
	\frac{\max_{1 \leq i \leq n} (w_i(x) -1/n)^2  }{\sum_{i=1}^n (w_i(x) - 1/n)^2 } = \frac{\max_{1 \leq i \leq n}  \left( K_h(x-X_i) - \frac{1}{n} \sum_{j=1}^n K_h(x-X_j)  \right)^2}{\sum_{i=1}^n  \left( K_h(x-X_i) - \frac{1}{n} \sum_{j=1}^n K_h(x-X_j)  \right)^2} \convP 0.
	\end{align*}
	Using the given condition, the numerator is bounded by 
	\begin{align*}
	\max_{1 \leq i \leq n}  \left( K_h(x-X_i) - \frac{1}{n} \sum_{j=1}^n K_h(x-X_j)  \right)^2 \leq 4 h^{-D} \mathcal{K}^2.
	\end{align*}
	Whereas the denominator can be decomposed into two parts:
	\begin{align*}
	& \sum_{i=1}^n  \left( K_h(x-X_i) - \frac{1}{n} \sum_{j=1}^n K_h(x-X_j)  \right)^2 \\[.5em]
	= ~ & \sum_{i=1}^n K_h^2(x-X_i)  - 2n \left(  \frac{1}{n} \sum_{j=1}^n K_h(x-X_j)  \right)^2
	\end{align*}
	Based on the usual bias-variance decomposition of the kernel density estimation \citep{wasserman2006all}, each part can be approximated as
	\begin{align*}
	& \frac{1}{nh^D}\sum_{i=1}^n K^2 \left(\frac{x-X_i}{h} \right) = f(x) \int K^2(u)du + O(h) + O_P \left( \frac{1}{\sqrt{nh^D}} \right)  \\[.5em]
	& \frac{1}{nh^D}\sum_{i=1}^n K \left(\frac{x-X_i}{h} \right) = f(x) + O(h^2)  + O_P \left( \frac{1}{\sqrt{nh^D}}  \right).
	\end{align*}
	Now, the sufficient condition can be further bounded by
	\begin{align} \nonumber
	& \frac{\max_{1 \leq i \leq n}  \left( K_h(x-X_i) - \frac{1}{n} \sum_{j=1}^n K_h(x-X_j)  \right)^2}{\sum_{i=1}^n  \left( K_h(x-X_i) - \frac{1}{n} \sum_{j=1}^n K_h(x-X_j)  \right)^2} \\[.5em] \nonumber
	\leq ~ & \frac{4 h^{-D} \mathcal{K}^2}{\frac{1}{h^{2D}}\sum_{i=1}^n K^2 \left(\frac{x - X_i}{h} \right) - 2n \left( \frac{1}{nh^D} \sum_{j=1}^n K\left( \frac{x-X_j}{h}\right)  \right)^2}  \\[.5em]
	= ~ & \frac{4 n^{-1}\mathcal{K}^2}{\frac{1}{nh^{D}}\sum_{i=1}^n K^2 \left(\frac{x - X_i}{h} \right) - 2h^D \left( \frac{1}{nh^D} \sum_{j=1}^n K\left( \frac{x-X_j}{h}\right)  \right)^2}. \label{Eq: Sufficient condition}
	\end{align}
	Then using the previous approximations, the denominator becomes
	\begin{align*}
	& ~ \frac{1}{nh^{D}}\sum_{i=1}^n K^2 \left(\frac{x - X_i}{h} \right) - 2h^D \left( \frac{1}{nh^D} \sum_{j=1}^n K\left( \frac{x-X_j}{h}\right)  \right)^2 \\[.5em]
	= ~ & f(x) \int K^2(u)du + O(h) + O_P \left( \frac{1}{\sqrt{nh^D}} \right) - 2h^D \left(  f(x) + O(h^2)  + O_P \left( \frac{1}{\sqrt{nh^D}}  \right)    \right)^2 \\[.5em]
	= ~ & \underbrace{f(x) \int K^2(u)du}_{>0 ~ \text{by the assumption}} + o_P(1).
	\end{align*}
	Hence (\ref{Eq: Sufficient condition}) converges to zero in probability and the result follows.
\end{proof}

\section{Diffusion Maps} \label{Sec: Diffusion Map}
Dimensionality reduction methods can be useful for visualizing and describing low-dimensional structures that are embedded in higher-dimensional spaces. In this section, we briefly describe diffusion maps \citep{coifman2005geometric, coifman2006diffusion} and the particular version that we use to visualize the results of our local two-sample test.

As a starting point for constructing a diffusion map, one first defines a weight that reflects the local similarity of two points $x_i$ and $x_j$ in $\mathcal{X} =\{x_1,\ldots, x_n \}$. A common choice is the Gaussian kernel 
\begin{align} \label{Eq: weight}
w(x_i,x_j) = \exp \left( - \frac{s(x_i,x_j)^2}{\epsilon}  \right),
\end{align}
where $s(x_i,x_j)$ represents (for example, the Euclidean) distance between the points. These weights are used to build a Markov random walk on the data with the transition probability from $x_i$ to $x_j$ defined as
\begin{align*}
p(x_i,x_j) = \frac{w(x_i,x_j)}{\sum_{k \in \Omega} w(x_i,x_k) } .
\end{align*}
The one-step transition probabilities are stored in an $n \times n$ matrix denoted by $\mathbf{P}$, and then usually propagated by a $t$-step Markov random walk with transition probabilities $\mathbf{P}^t$. 
Instead of choosing a fixed time parameter $t$, however, we here combine diffusions at all times \citep{coifman2005geometric} and define an averaged diffusion map\footnote{This is also the default option of the function {\tt diffuse()} in the {\tt R} package {\tt diffusionMap.}}  according to
\begin{align*}
\Psi_{\text{av}} :  x \mapsto \left[ \left(\frac{\lambda_1}{1-\lambda_1} \right) \psi_1(x), \left( \frac{\lambda_2}{1-\lambda_2} \right) \psi_2(x), \ldots, \left( \frac{\lambda_m}{1-\lambda_m} \right) \psi_m(x)   \right],
\end{align*}
where $\lambda_i$ and $\psi_i$, respectively, represent the first $m$th eigenvalues and the corresponding right eigenvectors of $\mathbf{P}$.

In our application for galaxy morphologies, we also use a generalization of the weight in (\ref{Eq: weight}) proposed by \cite{zelnik2005self} for spectral clustering. In their paper, the authors show that a data-driven varying bandwidth leads to more meaningful clustering results for data with multiple scales and propose the weight 
\begin{align*}
\widehat{w}(x_i,x_j) = \exp \left( - \frac{s(x_i,x_j)^2}{\sigma_i \sigma_j }  \right),
\end{align*}
where $\sigma_{i(j)}$ is the distance between $x_{i(j)}$ and its $k$th neighbor. For our visualization purposes, we choose $m=2$ and  $k=50$,  but a range of other values give similar results.

\end{document}